\newcommand*{\ATLASLATEXPATH}{}
\documentclass[PAPER, cernpreprint=true, texlive=2016, USenglish]{\ATLASLATEXPATH atlasdoc}
 
\usepackage[subcaption=true]{\ATLASLATEXPATH atlaspackage}
\usepackage{\ATLASLATEXPATH atlasbiblatex}
 
\usepackage{\ATLASLATEXPATH atlasphysics}
 
\protected\def\myphantom#1{\vphantom{#1}}

\newcommand{\DeltaTheta}{$\Delta_{\Theta(r)}$}
\newcommand{\RP}{$R_{P(r)}$}
\newcommand{\RTheta}{$R_{\Theta(r)}$}
\newcommand{\DeltaDptr}{\mbox{$\Delta\Dptr$}}
\newcommand{\Dptr}{\mbox{$D( \pT, r)$}}
\newcommand{\RDptr}{\mbox{$R_{D( \pT, r)}$}}

\newcommand{\PbPb}{\mbox{Pb+Pb}}
\newcommand{\pbpb}{\mbox{Pb+Pb}}

\newcommand{\Et}{\mbox{$E_{\mathrm{T}}$}}

\newcommand{\Ncoll}{\mbox{$N_{\mathrm{coll}}$}}

\newcommand{\Njet}{\mbox{$N_{\mathrm{jet}}$}}
\newcommand{\Nch}{\mbox{$n_{\mathrm{ch}}$}}

\newcommand{\nchmeas}{\mbox{$n_{\mathrm{ch}}^{\myphantom{\mathrm{j}}\mathrm{meas}}$}}
\newcommand{\nchsub}{\mbox{$n_{\mathrm{ch}}^{\myphantom{\mathrm{j}}\mathrm{sub}}$}}
\newcommand{\nchunf}{\mbox{$n_{\mathrm{ch}}^{\mathrm{unfolded}}$}}
\newcommand{\nch}{\mbox{$n_{\mathrm{ch}}$}}
\newcommand{\Npart}{\mbox{$N_{\mathrm{part}}$}}

\newcommand{\sqrtsnn}{\mbox{$\sqrt{s_{_\text{NN}}}$}}
\newcommand{\sqrts}{\mbox{$\sqrt{s}$}}

\newcommand{\pythiaeight}{\textsc{Pythia}8}

\newcommand{\powheg}{\textsc{Powheg}}

\newcommand{\RFour}{\mbox{$R = 0.4$}}

\newcommand{\TAA}{\mbox{$T_{\mathrm{AA}}$}}

\newcommand{\ETfcal}{\mbox{$\Sigma E_{\mathrm{T}}^{\mathrm{FCal}}$}}

\newcommand{\etajet}{\mbox{$\eta^{\mathrm{jet}}$}}
\newcommand{\phijet}{\mbox{$\phi^{\mathrm{jet}}$}}
\newcommand{\pTjet}{\mbox{$p_{{\mathrm{T}}}$}}

\newcommand{\pttrk}{\mbox{$p_{\mathrm{T}}^{\myphantom{\mathrm{j}}\mathrm{ch}}$}}

\newcommand{\pTtrue}{\mbox{$p_{\mathrm{T}}^{\mathrm{truth}}$}}
\newcommand{\etatrue}{\mbox{$\eta^{\mathrm{truth}}$}}

\newcommand{\ETjet}{\mbox{$\ET^{\mathrm{jet}}$}}

\newcommand{\ptjet}{\mbox{$p_{\mathrm{T}}^{\mathrm{jet}}$}}
\newcommand{\etatrk}{\mbox{$\eta^{\mathrm{ch}}$}}

\newcommand{\yjet}{\mbox{$y^{\mathrm{jet}}$}}

\newcommand{\rvar}{\mbox{$r$}}
\renewcommand{\_}{{\tt \char`\_}}

\graphicspath{{logos/}{figures/}}
 
\AtlasTitle{Measurement of angular and momentum distributions of charged particles within and around jets in Pb+Pb and $pp$ collisions at $\sqrt{s_{\mathrm{NN}}} = 5.02$~TeV with the ATLAS detector}

\AtlasAbstract{
Studies of the fragmentation of jets into charged particles in heavy-ion collisions can provide information about the mechanism of jet-quenching by the hot and dense QCD matter created in such collisions, the quark--gluon plasma.
This paper presents a measurement of the angular distribution of charged particles around the jet axis in $\sqrt{s_{\mathrm{NN}}}=$~5.02~TeV Pb+Pb and \textit{pp} collisions, using the ATLAS detector at the LHC.
The Pb+Pb and \textit{pp} data sets have integrated luminosities of 0.49~nb$^{-1}$ and 25~pb$^{-1}$, respectively.
The measurement is performed for jets reconstructed with the anti-$k_{t}$ algorithm with radius parameter $R = $~0.4 and is extended to an angular distance of $r= 0.8$ from the jet axis.
Results are presented as a function of Pb+Pb collision centrality and distance from the jet axis for charged particles with transverse momenta in the \mbox{1--63 GeV} range, matched to jets with transverse momenta in the  \mbox{126--316 GeV} range and an absolute value of jet rapidity of less than 1.7.
Modifications to the measured distributions are quantified by taking a ratio to the measurements in \textit{pp} collisions.
Yields of charged particles with transverse momenta below 4~GeV are observed to be increasingly enhanced as a function of angular distance from the jet axis, reaching a maximum at $r=0.6$.
Charged particles with transverse momenta above \mbox{4 GeV} have an enhanced yield in Pb+Pb collisions in the jet core for angular distances up to $r = 0.05$ from the jet axis, with a suppression at larger distances.
}
 
\author{The ATLAS Collaboration}
 
\AtlasRefCode{HION-2018-03}

\PreprintIdNumber{CERN-EP-2019-158}

\AtlasJournalRef{Phys.\ Rev.\ C. 100 (2019) 064901}
\AtlasDOI{10.1103/PhysRevC.100.064901}

\AtlasCoverSupportingNote{INT note on CDS}{https://cds.cern.ch/record/2304504}
 
\AtlasCoverCommentsDeadline{July 25th 2019}
 
\AtlasCoverAnalysisTeam{Akshat Puri, Anne Sickles, Martin Rybar}
 
\AtlasCoverEdBoardMember{Nachman, Benjamin  (LBL) }
\AtlasCoverEdBoardMember{Iwona, Grabowska-Bold (Krakow AGH-UST) }
\AtlasCoverEdBoardMember{Mario, Martinez  Perez (Barcelona) (chair)}
 
\AtlasCoverEgroupEditors{atlas-HION-2018-03-editors@cern.ch}
 
\AtlasCoverEgroupEdBoard{atlas-HION-2018-03-editorial-board@cern.ch}

\hypersetup{pdftitle={ATLAS document},pdfauthor={The ATLAS Collaboration}}
 
\begin{document}
 
\maketitle
 
\section{Introduction}
\label{sec:intro}
 
Ultra-relativistic nuclear collisions at the Large Hadron Collider (LHC) produce hot, dense matter called a quark--gluon plasma, QGP (see Refs.~\cite{Roland:2014jsa,Busza:2018rrf} for recent reviews).
Jets from hard-scattering processes in these collisions traverse and interact with the QGP, losing energy via a process called jet-quenching.
The rates and characteristics of these jets in heavy-ion collisions can be compared with the same quantities in \pp\ collisions, where the production of a QGP is not expected.
This comparison can provide information about the properties of the QGP and how it interacts with partons from the hard scatter.
 
Jets with large transverse momenta, \ptjet, in central lead--lead (\pbpb) collisions at the LHC are measured at approximately half the rates in \pp\ collisions when the nuclear overlap function of \pbpb\ collisions is taken into account~\cite{Abelev:2013kqa,Aad:2014bxa,Adam:2015ewa,Khachatryan:2016jfl, 2019108}.
Similarly, back-to-back dijet~\cite{Aad:2010bu,Chatrchyan:2011sx,Aaboud:2017eww} and photon--jet pairs~\cite{Chatrchyan:2012gt,Aaboud:2018anc} are observed to have less-balanced transverse momenta in \pbpb\ collisions compared to \pp\ collisions.
These observations suggest that some of the energy from the hard-scattered parton may be transferred outside of the jet through its interaction with the QGP medium.
 
Complementary measurements look at how the structure of jets in \pbpb\ collisions is modified relative to that in \pp\ collisions.
Previous measurements have shown a broadening of jets in \pbpb~\cite{Aad:2011sc, Acharya:2018uvf, Chatrchyan:2012mec, Chatrchyan:2013kwa}, as well as an excess of low- and high-momentum charged-particles and a depletion of intermediate-momentum charged-particles associated with these jets~\cite{Aad:2014wha,Chatrchyan:2014ava,Aaboud:2017bzv,Aaboud:2018hpb}.
Particles carrying a large fraction of the jet momentum are generally closely aligned with the jet axis, whereas low-momentum particles are observed to have a much broader angular distribution extending outside the jet~\cite{Chatrchyan:2011sx,Khachatryan:2015lha,Khachatryan:2016tfj,Sirunyan:2018jqr, Acharya:2019ssy}.
These observations suggest that the energy lost via jet-quenching is being transferred to soft particles around the jet axis via soft gluon emission~\cite{Vitev:2008rz,Ovanesyan:2011xy,Blaizot:2014ula,Qin:2015srf,Escobedo:2016jbm,Casalderrey-Solana:2016jvj,Tachibana:2017syd}.
Measurements of yields of these particles as a function of transverse momentum, \pt, and angular distance between the particle and the jet axis have the potential to provide further insight into the structure of jets in the QGP, as well as provide information about how the medium is affected by the presence of the jet.

This paper presents charged-particle \pt\ distributions at a distance \rvar\ around the jet axis\footnote{ATLAS uses a right-handed coordinate system with its origin at the nominal interaction point (IP) in the center of the detector, and the $z$-axis along the beam pipe.
The $x$-axis points from the IP to the center of the LHC ring, and the $y$-axis points upward.
Cylindrical coordinates $(r,\phi)$ are used in the transverse plane, $\phi$ being the azimuthal angle around the $z$-axis.
The pseudorapidity is defined in terms of the polar angle $\theta$ as $\eta=-\ln\tan(\theta/2)$.
The rapidity is defined as $y = 0.5\text{ln}[(E + p_z)/(E-p_z)]$ where $E$ and $p_z$ are the energy and $z$-component of the momentum along the beam direction respectively.
Transverse momentum and transverse energy are defined as $\pt = p \sin\theta$ and $\Et = E \sin\theta$, respectively.
The angular distance between two objects, $r$, with relative differences $\Delta \eta$ and $\Delta \phi$ in pseudorapidity and azimuth, respectively, is given by $\sqrt{(\Delta \eta )^2 + (\Delta \phi)^2}$.}
that have been corrected for detector effects.
The measured yields are defined as:
 
\begin{align*}
\Dptr = \frac{1}{N_{\mathrm{jet}}} \frac{1}{2\pi r dr} \frac{d n_{\mathrm{ch}} (\pt, r)}{d \pt},
\end{align*}
where $N_{\mathrm{jet}}$ is the number of jets in consideration and $n_{\mathrm{ch}}(\pt, r)$ is the number of charged particles with a given \pt\ at a distance \rvar\ from the jet axis.
The ratios of the charged-particle yields measured in \pbpb\ and \pp\ collisions,
 
\begin{align*}
\RDptr = \frac{\Dptr_\mathrm{Pb+Pb}}{\Dptr_{\pp}},
\end{align*}
quantify the modifications of the yields due to the QGP medium.
Furthermore, the differences between the \Dptr\ distributions in \pbpb\ and \pp\ collisions,
 
\begin{align*}
\Delta \Dptr = \Dptr_\mathrm{Pb+Pb} - \Dptr_{pp},
\end{align*}
allow the absolute differences in charged-particle yields between the two collision systems to be measured.

\section{ATLAS detector}
\label{sec:setup}
 
The measurements presented here are performed using the ATLAS \cite{Aad:2008zzm} calorimeter, inner detector, trigger, and data acquisition systems.

The calorimeter system consists of a sampling liquid-argon (LAr) electromagnetic (EM) calorimeter covering $|\eta|<3.2$, a steel--scintillator sampling hadronic calorimeter covering $|\eta| <1.7$, LAr hadronic calorimeters covering $1.5 < |\eta| < 3.2$, and two LAr forward calorimeters~(FCal) covering $3.1 < |\eta| < 4.9$.

The EM calorimeters are segmented longitudinally in shower depth into three layers with an additional presampler layer.
They have segmentation that varies with layer and pseudorapidity.
The hadronic calorimeters have three sampling layers longitudinal in shower depth.

The inner detector measures charged particles within the pseudorapidity interval $|\eta|<2.5$ using a combination of silicon pixel detectors, silicon microstrip detectors (SCT), and a straw-tube transition radiation tracker (TRT), all immersed in a 2~T axial magnetic field.
Each of the three detectors is composed of a barrel and two symmetric endcap sections.
The pixel detector is composed of four layers, including the insertable B-layer~\cite{ATLAS-TDR-2010-19, Abbott:2018ikt} added in 2014.
The SCT barrel section contains four layers of modules with sensors on both sides, and each endcap consists of nine layers of double-sided modules with radial strips.
The TRT contains layers of staggered straws interleaved with the transition radiation material.

The zero-degree calorimeters (ZDCs) are located symmetrically at $z = \pm140$~m and cover $|\eta| > 8.3$.
The ZDCs use tungsten plates as absorbers, and quartz rods sandwiched between the tungsten plates as the active medium.
In \PbPb\ collisions the ZDCs primarily measure `spectator' neutrons.
These are neutrons that do not interact hadronically when the incident nuclei collide.
A ZDC coincidence trigger is implemented by requiring the pulse height from both ZDCs to be above a threshold that accepts the signal corresponding to the energy deposition from a single neutron.

This analysis uses the same trigger setup used in Ref.~\cite{Aaboud:2018hpb} and is briefly described below.
A two-level trigger system was used to select the \PbPb\ and \pp\ collisions.
The first level, L1, is based on custom electronics, while the second level, the high-level trigger (HLT), is based on software \cite{Aaboud:2016leb}.
Minimum-bias~(MB) events were recorded using a logical OR of two triggers: 1) a total-energy L1 trigger selecting more-central collisions; 2) a ZDC coincidence trigger at L1 and a veto on the total-energy trigger, with the additional requirement of at least one track in the HLT, selecting peripheral collisions.
The total-energy trigger required the total transverse energy measured in the calorimeter system to be greater than 50~\GeV.
Jet events were selected by the HLT, seeded by a jet identified by the L1 jet trigger in \pp\ collisions or by the total-energy trigger with a threshold of 50~\GeV\ in \PbPb\ collisions.
The L1 jet trigger utilized in \pp\ collisions required a jet with transverse momentum greater than 20~\GeV.
The HLT jet trigger uses a jet reconstruction procedure similar to that in the offline analysis as discussed in Section~\ref{sec:reconstruction}.
It selected events containing jets with a transverse energy of at least 75~\GeV\ in \PbPb\ collisions and at least 85~\GeV\ in \pp\ collisions.
The measurement is performed in the jet transverse momentum range where the trigger efficiencies are greater than 99\%.

\section{Data sets and event selection}
\label{sec:data}
 
The \PbPb\ and \pp\ data used in this analysis were recorded in 2015.
The data samples consist of 25~pb$^{-1}$ of \sqrts~=~5.02~\TeV\ \pp\ and 0.49~nb$^{-1}$ of \sqrtsnn~=~5.02~\TeV\ \pbpb\ data.
In both samples, events are required to have a reconstructed vertex within 150~mm of the nominal IP along the beam axis.
Events with multiple interactions in the same bunch crossing are referred to as pileup.
This is negligible in the \pbpb\ data, and the \pp\ data was collected in low-pileup mode.
The average number of interactions per bunch crossing in the \pp\ collisions ranged from 0.6 to 1.3.
Only events taken during stable beam conditions and satisfying detector and data-quality requirements that include the detector subsystems being in normal operating condition are considered.
 
In \pbpb\ collisions, the event centrality reflects the overlap area of the two colliding nuclei and is characterized by \ETfcal, the total transverse energy deposited in the FCal~\cite{Aaboud:2017tql}.
The six centrality intervals used in this analysis are defined according to successive percentiles of the \ETfcal\ distribution obtained in MB collisions, ordered from the most-central (highest \ETfcal) to the most-peripheral (lowest \ETfcal) collisions: 0--10\%, 10--20\%, 20--30\%, 30--40\%, 40--60\%, 60--80\%.
 
The \pp\ Monte Carlo (MC) sample consists of $1.8\times10^7$ 5.02 TeV hard-scattering dijet \pp\ events generated with \powheg{}+\pythiaeight\ \cite{Nason:2004rx,Frixione:2007vw, Alioli:2010xd, Alioli:2010xa, Sjostrand:2014zea} using the A14 tune of parameter values \cite{ATLAS2014021} and the NNPDF23LO PDF set \cite{Ball:2012cx}.
The \pbpb\ MC sample was generated by overlaying an additional sample of MB \pbpb\ data events onto a separate set of $1.8\times10^7$ 5.02~\TeV\ hard-scattering dijet \pp\ events generated with the same tune and PDFs as the \pp\ MC sample.
This `MC overlay' sample was reweighted on an event-by-event basis such that it has the same centrality distribution as the jet-triggered data sample.
Another sample of MB \pbpb\ events was generated using HIJING (version 1.38b) \cite{Wang:1991hta} and was only used to evaluate the track reconstruction performance.
The detector response in all MC samples was simulated using \mbox{\textsc{Geant4} \cite{Agostinelli:2002hh,Aad:2010ah}}.
These MC samples are used to evaluate the performance of the detector and analysis procedure and correct the measured distributions for detector effects.

\section{Jet and track selection}
\label{sec:reconstruction}
 
The jet reconstruction procedure is identical to that used in Ref.~\cite{2019108}.
The \antikt\ algorithm \cite{Cacciari:2008qp, Fastjet} is first run in four-momentum recombination mode on $\Delta \eta \times \Delta \phi = 0.1\times 0.1$  calorimeter towers with two \antikt\ radius parameter values ($R=$~0.2 and $R=$~0.4).
The energies in the towers are obtained by summing the energies of calorimeter cells at the electromagnetic energy scale.
Then, an iterative procedure is used to estimate the $\eta$-dependent underlying event (UE)  transverse energy density, while excluding the regions populated by jets.
The estimate of the UE contribution is performed on an event-by-event basis.
Furthermore, the background in \pbpb\ collisions is modulated to account for the azimuthal anisotropy in particle production~\cite{ATLAS:2012at}.
The modulation includes the contributions of the second-, third-, and fourth-order azimuthal anisotropy harmonics.
The UE transverse energy is subtracted from calorimeter towers included in the jet and the four-momentum of the jet is updated accordingly.
Then, a jet $\eta$- and \pT-dependent  correction factor to the \ptjet\ derived from the \pp\ MC sample is applied to correct for the calorimeter energy response~\cite{Aaboud:2017jcu}.
The same calibration factors are applied in both the \pp\ and \pbpb\ collisions.
An additional correction based on \textit{in situ} studies of jets recoiling against photons, $Z$ bosons, and jets in other regions of the calorimeter is applied~\cite{2019167}.
The same jet reconstruction procedure without the azimuthal modulation of the UE is also applied to $pp$ collisions.
The UE subtraction in $pp$ collisions removes the pileup contribution to the jet.
In this analysis, jets are required to have \ptjet\ in the 126--316~\GeV\ range, with rapidity  $|\yjet|<$~1.7.
The \ptjet\ requirement is chosen in order to exclude the contribution of `UE jets' generated by fluctuations in the UE, while the rapidity requirement is based on the acceptance of the tracking system.
To prevent nearby jets from distorting the measurement of \Dptr\ distributions, jets are rejected if there is a neighboring jet with higher \ptjet\ within an angular distance of $\Delta R = 1.0$.
This isolation requirement removes approximately 0.01\% of jets.
Additionally, generator-level jets are reconstructed by applying the \antikt\ algorithm with radius parameter $R=0.4$ to stable final-state particles from MC generators.
These particles are required to have a lifetime of $\tau>0.3\times 10^{-10}$ s and do not include muons, neutrinos, and particles from pileup activity.
 
Charged-particle tracks in \pbpb\ collisions are reconstructed from hits in the inner detector using the track reconstruction algorithm that was optimized for the high hit density in heavy-ion collisions~\cite{Aaboud:2017all}.
Tracks used in this analysis have $|\eta| < 2.5$ and are required to have at least 9 (11) total hits in the silicon detectors for charged particles with pseudorapidity \mbox{$\left|\etatrk\right| \leq 1.65 \left( \left|\etatrk\right| > 1.65 \right)$}.
At least one hit is required in one of the two innermost pixel layers.
If the track trajectory passes through an active module in the innermost layer, then a hit in this layer is required.
Additionally, a track must have no more than two holes in the pixel and SCT detectors together, where a hole is defined by the absence of a hit predicted by the track trajectory.
All charged-particle tracks used in this analysis are required to have reconstructed transverse momentum $\pttrk > 1.0 $~\GeV.
In order to suppress the contribution from secondary particles,\footnote{Primary particles are defined as particles with a mean lifetime $\tau>0.3\times 10^{-10}$ s either directly produced in \pp\ interactions or from subsequent decays of particles with a shorter lifetime.
All other particles are considered to be secondary.} the distance of closest approach of the track to the primary vertex is required to be less than a value that varies from 0.45~mm at $\pttrk=4$~\GeV\ to 0.2~mm at $\pttrk=20$~\GeV\ in the transverse plane and less than 1.0~mm in the longitudinal direction.
The primary vertex is determined using vertex finding and fitting algorithms described in Ref.~\cite{ATLAS-CONF-2010-069}.
The precision is better than $20\mu m$ in the transverse plane in peripheral collisions and improves inversely proportional to the square root of the number of reconstructed primary tracks in central collisions.

The efficiency, $\varepsilon$, for reconstructing charged particles in \PbPb\ and \pp\ collisions is determined using the MC samples described above.
It is evaluated as a function of the generator-level primary particle transverse momentum, \pTtrue, and pseudorapidity, \etatrue, by matching tracks to generator-level primary particles~\cite{Aad:2010ah}.
In both collision systems the efficiency increases slowly with \pTtrue\ and is seen to be independent of \ptjet\ in the measurement's phase space.
For \pbpb\ collisions, the efficiency for $|\eta| < 0.3$  is $\sim 80\%$ at 1~\GeV\ and rises to $\sim 85\%$ at 10~\GeV.
For $1.0 < |\eta| < 2.0$, the efficiency is $\sim 67\%$ to $\sim 72\%$  over the same \pt\ range, with the variation in efficiency between the most-central and most-peripheral \pbpb\ collisions being approximately 3\% in both $\eta$ ranges.
For \pp\ collisions, the efficiency for $|\eta| < 0.3$ is $\sim85\%$ at 1~\GeV, and rises to $\sim 88 \%$ at 10~\GeV, remaining relatively constant thereafter.
For $1.0 < |\eta| < 2.0$, the efficiency is $\sim 82\%$ to $\sim 86\%$ over the same \pt\ range.
Further details about the tracking efficiency can be found in Ref.~\cite{Aaboud:2017bzv}.
 
The contribution of reconstructed tracks that cannot be matched to a generated primary particle in the \pp\ MC samples is less than 2\% in the entire \pttrk\ range under study in both the \pp\ and \pbpb\ collisions.
This contribution includes fakes and secondaries, where fakes originate from randomly matched hits in the detector layers that do not correspond to the passage of charged particles.
Both of these contributions are corrected for as described in the next section.

\section{Analysis procedure}
\label{sec:analysis}
 
The analysis procedure is similar to that in Ref.~\cite{Aaboud:2018hpb}, except it is also performed differentially in \rvar.
Measured tracks are considered to be associated with a reconstructed jet if their angular distance from the jet axis is less than $0.8$.
Their multiplicity distribution is given by:
 
\begin{align*}
\dfrac{d^{2} \nchmeas ( \pttrk, r)}{d \pttrk d r} = \frac{1}{\varepsilon(\pttrk,\etatrk)} \frac{\Delta \Nch ( \pttrk, r)}{ \Delta\pttrk \Delta r}
\end{align*}
where $\Delta \Nch (\pttrk, r)$ represents the number of tracks within a given $\Delta \pttrk$ and $\Delta r$ range.
The efficiency correction is applied as a $1/\varepsilon(\pttrk,\etatrk)$ weight on a track-by-track basis, assuming $\pttrk = \pTtrue$.
While that assumption is not strictly valid, the efficiency varies sufficiently slowly with $\pTtrue$ that the error introduced by this assumption is less than 1\%, and is further corrected for by the Bayesian unfolding procedure described later in this section.
 
The measured track yields need to be corrected for the contributions from the UE, fake tracks, and secondary particles.
In \pp\ collisions, the UE contribution from hard scatterings not associated with jet production is negligible.
The contributions from fake tracks and secondary charged particles are estimated from MC samples and subtracted.
This procedure is similar to that applied in previous measurements~\cite{Aaboud:2017tke,Aaboud:2018hpb}.
 
\begin{figure}
\includegraphics[width=1.\textwidth]{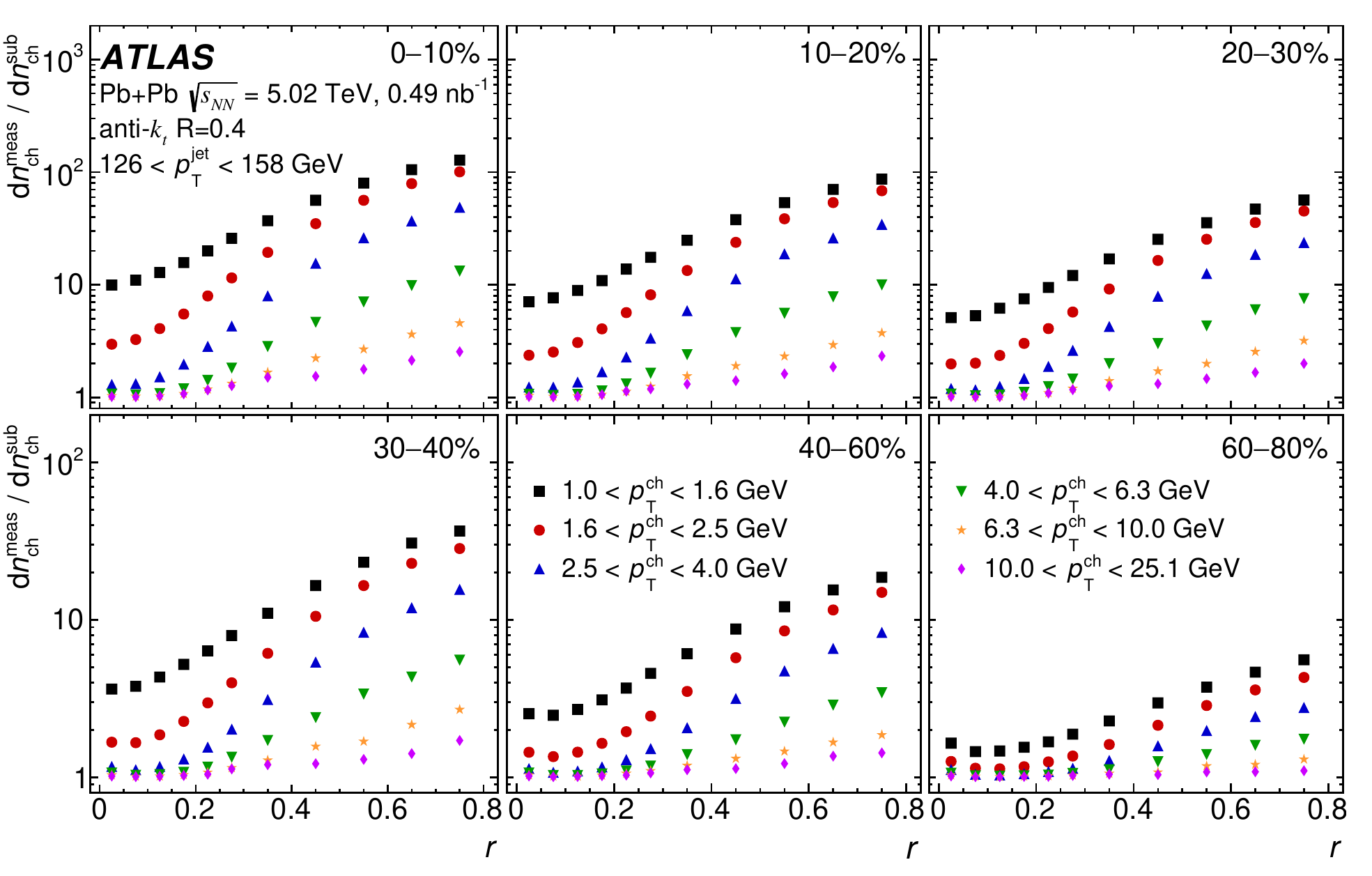}
\caption{Ratio of the measured charged-particle distributions to those after the subtraction of the UE, fake tracks, and secondaries as a function of \rvar\ for different \pttrk\ intervals for \ptjet\ in the range 126--158~\GeV.
The different panels represent the six centrality selections.}
\label{fig:UEsize}
\end{figure}

For \pbpb\ collisions, contributions from the UE, fake tracks, and secondary particles are estimated together in a two-step process:
first, the MC overlay sample is used to generate \etajet\ vs.\ \phijet\ maps of the average number of charged particles in a given annulus around a reconstructed jet.
This is done for charged particles without an associated generated primary particle and as a function of \ptjet, \etajet, \phijet,  $\Delta\Psi_{\mathrm{jet}}$, \rvar, \pttrk, and centrality.
Here $\Delta\Psi_{\mathrm{jet}}$ is the azimuthal angle between the jet and the second-order event plane $\Psi_2$ and is given by $ \Delta\Psi_{\mathrm{jet}} = \phijet - \Psi_2$.\footnote{The second-order event plane angle $\Psi_2$ is determined on an event-by-event basis by a standard method using the $\phi$ variation of transverse energy in the FCal \cite{ATLAS:2012at}.}
In the second step, the \etajet\ vs.\ \phijet\ maps are used to construct the UE distribution in the data as a function of \ptjet, \etajet, \phijet, $\Delta\Psi_{\mathrm{jet}}$, and centrality.
This distribution includes fakes and tracks from secondary particles, and is given by \mbox{$d^2 \nch^{\mathrm{UE+Fake}}(\pttrk, r) / d \pttrk d r$}.
The yields decrease with decreasing collision centrality, increasing \pttrk, and increasing $\Delta\Psi_2$.
The subtracted distributions are then given by:
 
\begin{align*}
\frac{ d^2 \nchsub ( \pttrk, r) }{ d \pttrk d r } &=  \frac{ d^2 \nchmeas  ( \pttrk, r) }{ d \pttrk d r} -  \frac{ d^2 \nch^{\mathrm{UE+Fake}}( \pttrk, r)  }{ d \pttrk d r}.
\end{align*}
Figure~\ref{fig:UEsize} shows the ratio of the charged-particle distributions before and after the subtraction of the UE, fake tracks, and secondaries, as a function of \rvar\ for different \pttrk\ intervals and $126 < \ptjet < 158$~\GeV\ for six centrality selections.
The largest UE contribution is for 1.0~\GeV\ charged particles at large values of \rvar\ in central collisions, with the background-to-signal ratio being approximately 100 and slowly decreasing with increasing \ptjet.
It rapidly decreases for more-peripheral collisions, larger \pttrk, and smaller \rvar.
 
To remove the effects of bin migration due to the jet energy and track momentum resolution, the subtracted $d^2 \nchsub /d\pttrk d r$ distributions are corrected by a two-dimensional Bayesian unfolding~\cite{DAgostini:1994zf} in \pttrk\ and \ptjet\ as implemented in the RooUnfold package~\cite{Adye:2011gm}.
Two-dimensional unfolding is used because the calorimetric jet energy response depends on the fragmentation pattern of the jet~\cite{Aad:2011he}.
Four-dimensional response matrices are created from the \pp\ and \pbpb\ MC samples using the generator-level and reconstructed \ptjet and the generator-level and reconstructed charged-particle \pttrk.
They are corrected for tracking efficiencies and are evaluated in bins of \rvar\ and centrality.
The Bayesian procedure requires a choice in the number of iterations.
Additional iterations reduce the sensitivity to the choice of prior, but may amplify statistical fluctuations in the distributions.
After four iterations the charged-particle distributions are found to be stable within 2--4\% for both the \PbPb\ and \pp\ data.
A separate one-dimensional Bayesian unfolding is used to correct the measured \ptjet\ spectra that are used to normalize the unfolded charged-particle distributions.
The response matrices for both the one- and two-dimensional unfolding are reweighted such that the \Dptr\ and \ptjet\ distributions match the shapes of the corresponding quantities in the reconstructed data.
 
An independent bin-by-bin unfolding procedure is used to correct for migrations originating from the jet and track angular resolutions.
Two corresponding \Dptr\ distributions are evaluated in MC samples, one using generator-level jets and primary particles, and the other using reconstructed jets and charged particles with their reconstructed \pt\ replaced by generator-level transverse momentum, \pTtrue.
The ratio of these two MC distributions provides a correction factor which is then applied to the data.
These factors are at the level of approximately 5\% with variations up to 15\% for particles with $\pt > 4$~\GeV, particularly near the edge of the jet.
For 126--158~\GeV\ jets in \pbpb\ collisions, the full unfolding correction is 8--14\% (4--7\%) for 1~\GeV\ (6~\GeV) tracks, depending on centrality.

The final particle-level corrected distributions, normalized by the area of the annulus being studied, are defined as:
 
\begin{align*}
\Dptr = \frac{1}{N_\mathrm{jet}^\mathrm{unfolded}} \frac{1}{2 \pi r dr} \frac{d \nchunf ( \pttrk, r) }{d \pt},
\end{align*}
where $N_\mathrm{jet}^\mathrm{unfolded}$ is the unfolded number of jets in a given \ptjet\ interval, and \nchunf\ is the unfolded yield of charged particles with a given \pt\ matched to a jet with given \ptjet, at a distance \rvar.
 
The performance of the full analysis procedure is validated in the MC samples by comparing the fully corrected charged-particle distributions to the generator-level distributions.
Good recovery of the generator-level distributions (closure) is observed, with a variation of less than 4\% for charged particles with \mbox{$\pt < 10$ GeV} in both the \pp\ and \pbpb\ collision systems.
The non-closure is taken as an additional systematic uncertainty as discussed in Section~\ref{sec:systematics}.
One important effect is that adding or removing particles carrying a large fraction of the jet momentum near the edge of the jet can significantly alter its reconstructed momentum and direction; this instability contributes to the non-closure mentioned above for particles with $\pt > 10$~\GeV\ in jets with $\ptjet < 200$~\GeV.
Results are presented only where the non-closure in the \pp\ MC sample is less than 5\%.

\section{Systematic uncertainties}
\label{sec:systematics}
 
The following sources of systematic uncertainty are considered: the jet energy scale (JES), the jet energy resolution (JER), the sensitivity of the  unfolding to the prior, the UE contribution, the residual non-closure of the analysis procedure, and tracking-related uncertainties.
For each variation accounting for a source of systematic uncertainty, the \Dptr\ distributions along with their ratios and differences are re-evaluated.
The difference between the varied and nominal distributions is used as an estimate of the uncertainty.

The systematic uncertainty associated with the JES in \PbPb\ collisions is due to jets having a different structure and possibly a different detector response that is not modeled by the MC simulation.
It is composed of two parts: a centrality-independent baseline component and a centrality-dependent component.
Only the centrality-independent baseline component is used in \pp\ collisions; it is determined from \textit{in situ} studies of the calorimeter response~\cite{Aad:2011he,HIjesnote,Aaboud:2017jcu} and the relative energy scale difference between the jet reconstruction procedures in heavy-ion~\cite{HIjesnote} and \pp\ collisions~\cite{Aaboud:2017jcu}.
The centrality-dependent uncertainty reflects a modification of parton showers by the \PbPb\ environment.
It is evaluated by comparing the ratio of the transverse momentum of calorimeter jets to the vectorial sum of the transverse momentum of charged particles within these jet in data and MC events.
More details on this procedure can be found in Ref.~\cite{2019108}.
The size of the centrality-dependent uncertainty in the JES reaches 0.5\% in the most-central collisions.
Each component that contributes to the JES uncertainty is varied separately by $\pm1$ standard deviation for each interval in \ptjet, and the response matrix is recomputed accordingly.
The data are then unfolded with the modified matrices.
The resulting uncertainty from the JES increases with increasing charged-particle \pT\ at fixed \ptjet\ and decreases with increasing \ptjet, and is at the level of 2--4\%.

The uncertainty in the \Dptr\ distributions due to the JER is evaluated by repeating the unfolding procedure with modified response matrices, where an additional contribution is added to the resolution of the reconstructed \ptjet\ using a Gaussian smearing procedure.
The smearing factor is evaluated using an \textit{in situ} technique in 13~\TeV\ \pp\ data that involves studies of dijet energy balance~\cite{Aad:2012ag,JERConfNote}.
An additional uncertainty is included to account for differences between the tower-based heavy-ion jet reconstruction and that used in analyses of 13~\TeV\ \pp\ data \cite{Aaboud:2017jcu, HIjesnote}.
The resulting uncertainty from the JER is symmetrized to account for negative variations of the JER.
The size of the resulting uncertainty in the \Dptr\ distributions due to the JER typically reaches 4--5\% for the highest charged-particle \pT\ intervals and decreases to 2--3\% with decreasing charged-particle \pT\ at fixed \ptjet.

The uncertainties related to track reconstruction and selection originate from several sources.
Uncertainties related to the detector material description in simulation and the track transverse momentum resolution are obtained from studies in data and simulation described in Ref.~\cite{ATL-PHYS-PUB-2015-051}.
The sensitivity of the tracking efficiency to the description of the inactive material in the MC samples is evaluated by varying the material description.
This resulting uncertainty in the track reconstruction efficiency is between 0.5\% and 2\% in the track \pT\ range used in the analysis.
The systematic uncertainty associated with the rate of fakes and secondary tracks is 30\% in both collision systems~\cite{ATL-PHYS-PUB-2015-051}.
The contamination of the \nchsub\ from fakes and secondary tracks is approximately 20\% in the jet core for tracks with $1 < \pt < 1.6$ \GeV.
It rapidly decreases as a function of \pt\ and \rvar\ and is less than 2\% at larger distances from the jet axis.
The resulting uncertainty in the \Dptr\ distributions is at most $\sim7\%$ in the jet core and decreases as a function of \pt\ and \rvar.
An additional uncertainty takes into account a possible residual misalignment of the tracking detectors in \pp\ and \PbPb\ data-taking.
The resulting uncertainties in the \Dptr\ distributions are typically less than 0.1\%.
An additional uncertainty in the tracking efficiency due to the high local track density in the core of jets is 0.4\%~\cite{ATL-PHYS-PUB-2016-007} for all \ptjet\ ranges in this analysis.
The uncertainty due to the track selection is evaluated by repeating the analysis with an additional $3\sigma$ requirement on the significance of the distance of closest approach of the track to the primary vertex.
This uncertainty affects the track reconstruction efficiencies, track momentum resolution, and rate of fake tracks.
The resulting uncertainty typically varies between 1\% and 2\%.
Finally, the track-to-particle matching requirements are varied.
This variation affects the track reconstruction efficiency, track momentum resolution, and rate of fake tracks.
The resulting systematic uncertainty is $\leq0.1$\% for the \Dptr\ distributions.
All track-related systematic uncertainties are added in quadrature and presented as the total tracking uncertainty.

The systematic uncertainty associated with the UE subtraction has two components: the limited number of charged particles associated with a jet without a corresponding generator particle in the \pbpb\ MC overlay sample, and a comparison with the result of an alternative UE estimation using the cone method.
The cone method uses jet-triggered events to estimate the background and is adapted from Refs.~\cite{Aaboud:2018hpb, Aaboud:2017bzv}.
A regular grid of nine cones of size $\Delta R = 0.8$ is used to cover the inner-detector region.
Cones are excluded if the angular distance to any reconstructed jet in the event with $\ptjet > 90$~\GeV\ is less than 1.6 or if they contain a charged particle with \mbox{$\pt > 10$ \GeV}.
This exclusion reduces biases from any hard processes.
The resulting UE charged-particle yields $d n_{\mathrm{ch}}^{\mathrm{UE, Cone}} / d \pttrk$ are evaluated over the \mbox{1--10 \GeV} range as a function of \pttrk, \ptjet, centrality, and \rvar, and are subsequently averaged over all cones.
Both of these sources of uncertainty are combined as uncorrelated uncertainties.
The combined UE uncertainty in the \Dptr\ distributions is less than 10\% for $\rvar < 0.4$ and reaches a maximum of 40\% at the largest angular distances from the jet axis.
It is the dominant source of systematic uncertainty for low-\pt\ charged particles at large \rvar\ and decreases sharply with increasing charged-particle \pT.
In particular, the component from the limited sample size dominates in the most-central collisions, while the component from the alternative estimation method dominates elsewhere.

The systematic uncertainty in the unfolding procedure is estimated by generating response matrices from the MC distributions without the reweighting that matched the shapes of the charged-particle and jet distributions in data.
The 5--7\% difference between the nominal \Dptr\ distribution and that unfolded with the un-reweighted response matrix is taken as the systematic uncertainty.
 
An additional uncertainty to account for possible residual limitations in the analysis procedure is assigned by evaluating the non-closure of the unfolded distributions in simulations.
This is typically about 4\%.
 
The correlations between the various systematic components are considered in evaluating the \RDptr\ and $\Delta\Dptr$ distributions.
The unfolding and non-closure uncertainties are taken to be uncorrelated between \pp\ and \pbpb\ collisions, while all others are taken to be correlated.
For these, the \RDptr\ and $\Delta\Dptr$ distributions are re-evaluated by applying the variation to both collision systems; the resulting variations of the ratios from their central values are used as the correlated systematic uncertainty.

\begin{figure}
\includegraphics[width=1.\textwidth]{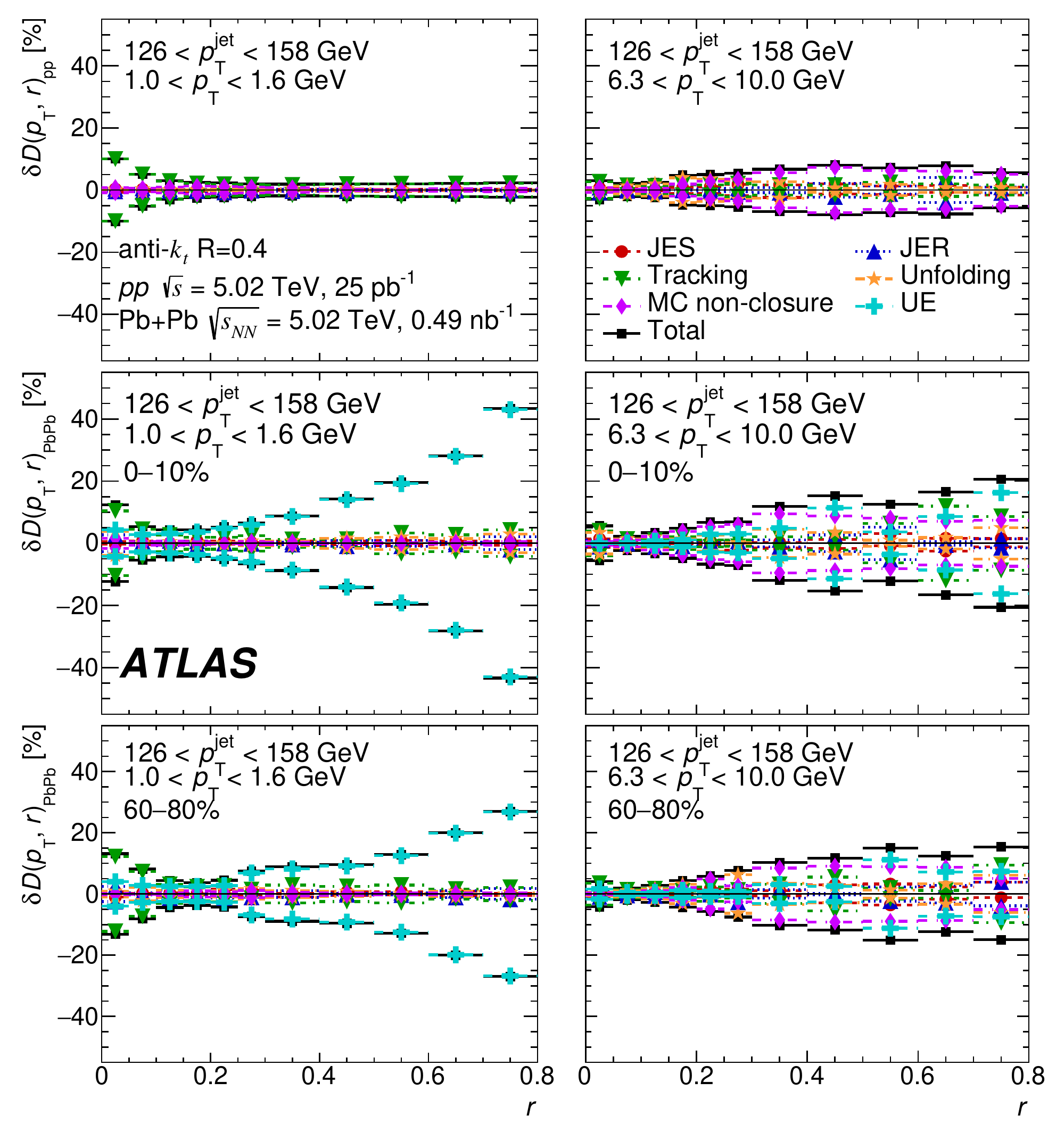}\hspace{0.0em}
\caption{Relative size of the systematic uncertainties for \Dptr\ distributions in \pp\ (top), central 0--10\% \pbpb\ (middle), and peripheral 60--80\% \pbpb\ (bottom) collisions for tracks with $1.0 < \pt < 1.6$~\GeV\ (left) and $6.3 < \pt < 10$~\GeV\ (right) in jets with $126 < \ptjet < 158$~\GeV.
The systematic uncertainties due to JES, JER, tracking, unfolding, MC non-closure, and UE contribution (for \pbpb) are shown along with the total systematic uncertainty from all sources.}
\label{fig:Systematics_Dpt}
\end{figure}

\begin{figure}
\includegraphics[width=1.\textwidth]{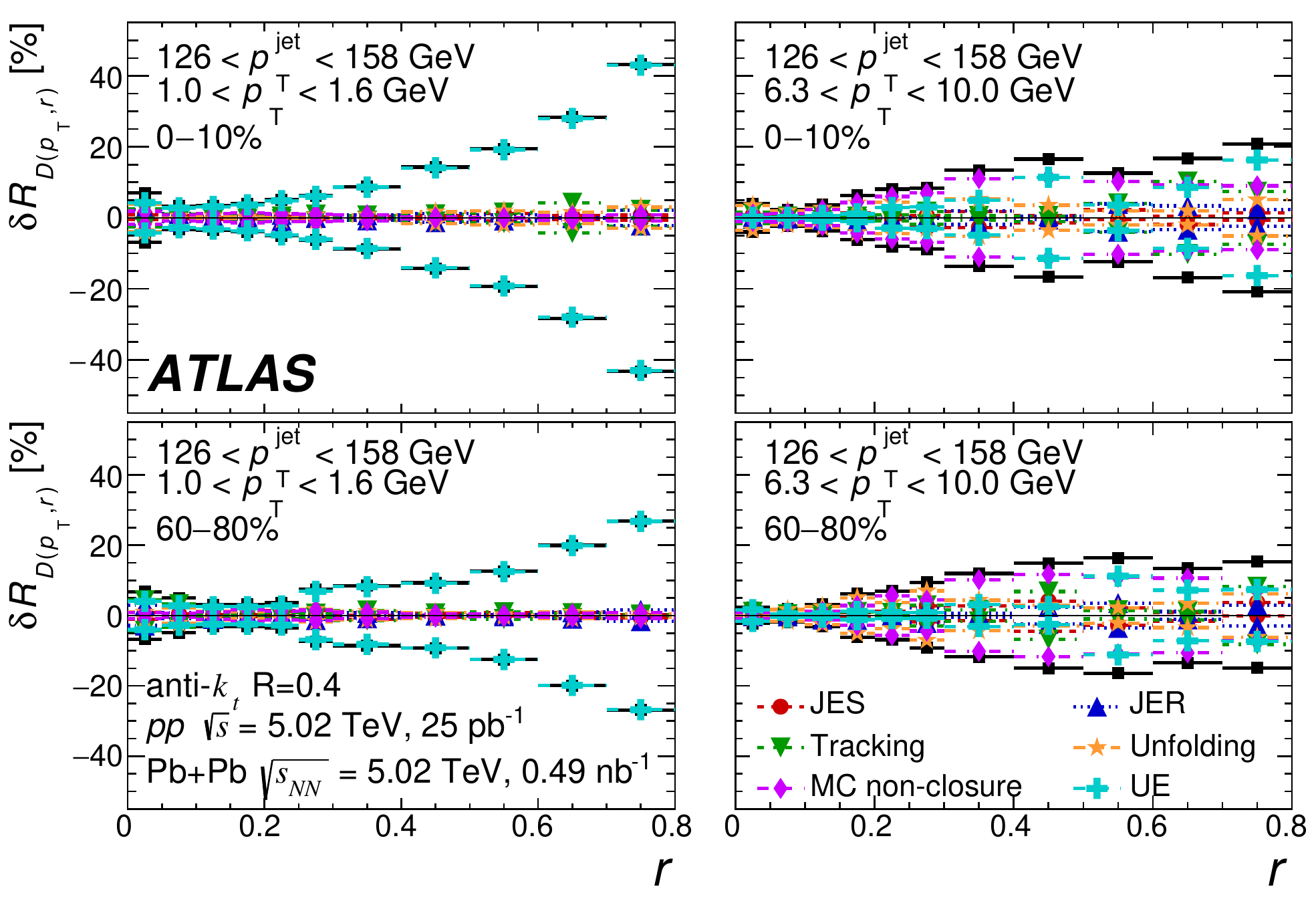}\hspace{0.0em}
\caption{Relative size of the systematic uncertainties for \RDptr\ distributions for 0--10\% (top) and 60--80\% (bottom) \pbpb\ collisions, for tracks with $1.0 < \pt < 1.6$~\GeV\ (left) and $6.3 < \pt < 10.0$~\GeV\ (right), in jets with $126 < \ptjet < 158$~\GeV.
The systematic uncertainties due to JES, JER, tracking, unfolding, MC non-closure, and UE contribution are shown along with the total systematic uncertainty from all sources.}
\label{fig:Systematics_RDpT}
\end{figure}

Examples of systematic uncertainties in the \Dptr\ distributions for jets in the 126--158~\GeV\ \ptjet\ range measured in \pp\ and \pbpb\ collision systems are shown in Figure~\ref{fig:Systematics_Dpt}.
The uncertainties in the \RDptr\ distributions are shown in Figure~\ref{fig:Systematics_RDpT}.
It can be seen that the dominant systematic uncertainty in the \pbpb\ and the \RDptr\ distributions is from the UE subtraction.
While it is less than 5\% for $r < 0.3$, it is approximately 40\% for charged particles with $\pt = 1$~\GeV\ at $r = 0.8$.
The uncertainties in the \pp\ system are smaller, with the dominant systematic uncertainty at low \pt\ being due to the tracking.
This is approximately 10\%  for $r < 0.1$ and decreases to less than 5\% at larger distances.

\section{Results}
\label{sec:results}
 
The \Dptr\ distributions are studied as a function of \ptjet\ for \pp\ data and \PbPb\ collisions with different centralities.
The interplay between the hot and dense matter and the parton shower is explored by evaluating the ratios and differences between the \Dptr\ distributions in \pbpb\ and \pp\ collisions.
Some selected moments of these distributions are also investigated.

\subsection{\Dptr\ distributions}
\label{sec:dptr}
\begin{figure}
\includegraphics[width=1.\textwidth]{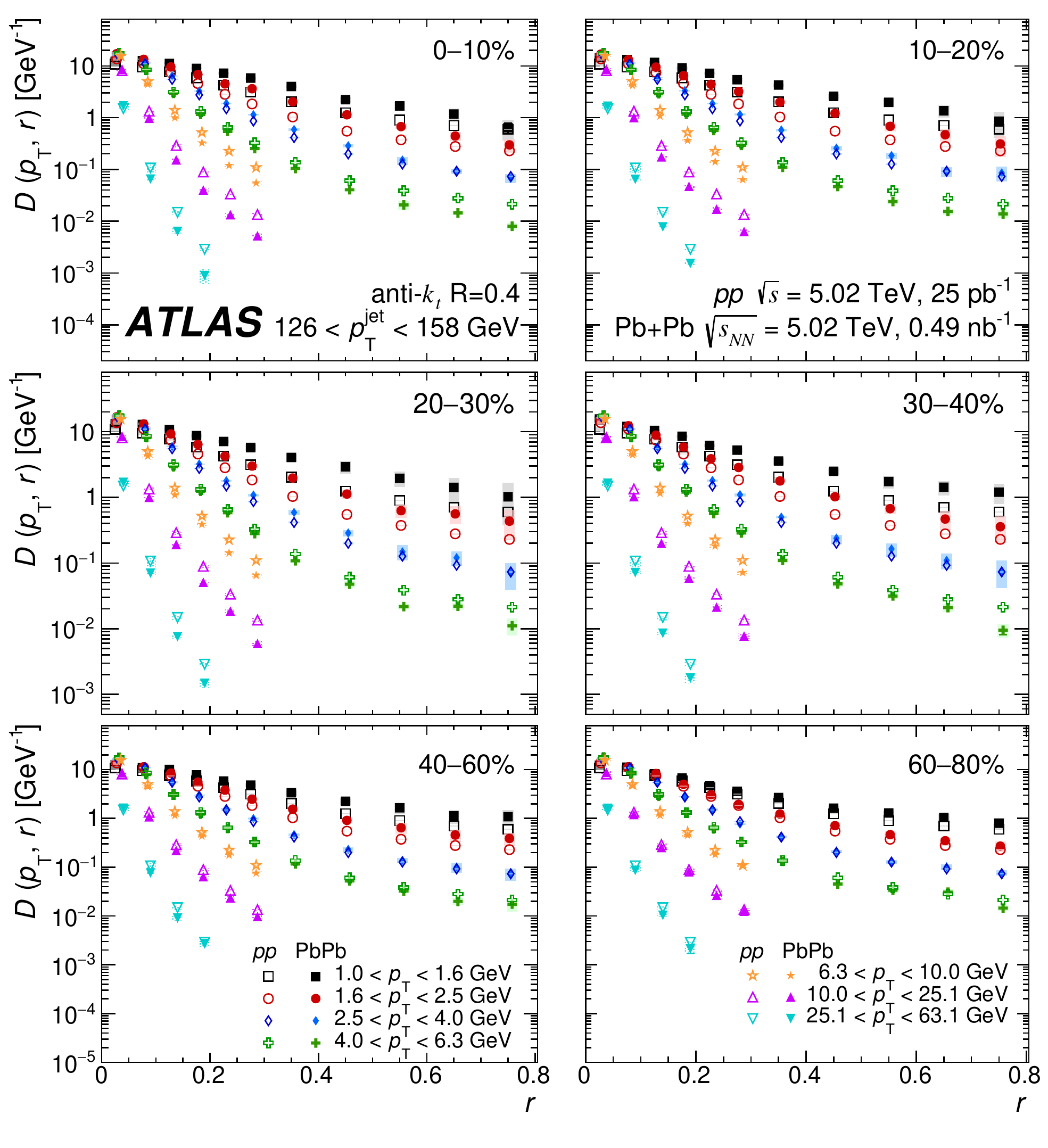}\hspace{0.0em}
\caption{The \Dptr\ distributions in \pp\ (open symbols) and \pbpb\ (closed symbols) as a function of angular distance $r$ for \ptjet\ of 126 to 158~\GeV.
The symbols represent different track \pt\ ranges, and each panel is a different centrality selection.
The vertical bars on the data points indicate statistical uncertainties, while the shaded boxes indicate systematic uncertainties.
The widths of the boxes are not indicative of the bin size, and the points are shifted horizontally for better visibility.
The distributions for $\pt > 6.3$~\GeV\ are restricted to smaller \rvar\ values as discussed in Section~\ref{sec:analysis}.}
\label{fig:dptr}
\end{figure}

The \Dptr\ distributions evaluated in \pp\ and \pbpb\ collisions for $126 < \ptjet < 158$~\GeV\ are shown in Figure~\ref{fig:dptr}.
These distributions decrease as a function of distance from the jet axis.
The rate of fall-off increases sharply for higher-\pt\ particles, with most of these being concentrated near the jet axis.
The distributions exhibit a difference in shape between \PbPb\ and \pp\ collisions, with the \pbpb\ distributions being broader at low \pt\ (\pt < 4~\GeV) and narrower at high \pt\ (\pt > 4~\GeV) in \mbox{0--10\%} central collisions.
This modification is centrality dependent and is smaller for peripheral \pbpb\ collisions.

\subsection{\RDptr\ distributions}
\label{sec:rdptr}
\begin{figure}
\includegraphics[width=1.\textwidth]{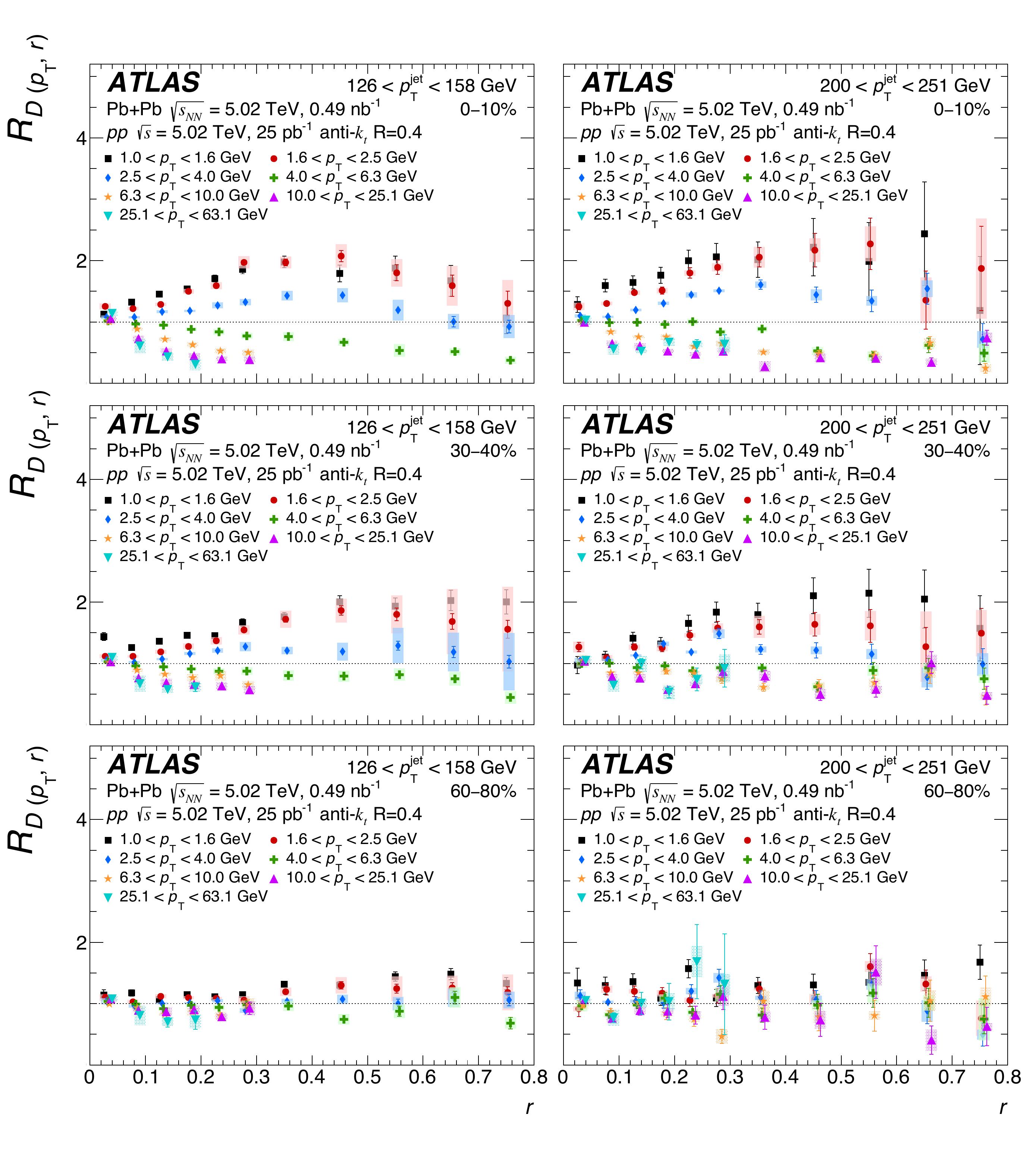}\hspace{0.0em}
\caption{Ratios of \Dptr\ distributions in \PbPb\ and \pp\ collisions as a function of angular distance $r$ for $126 < \ptjet < 158$~\GeV\ (left) and $200 < \ptjet < 251$~\GeV\ (right) for seven \pt\ selections.
Different centrality selections are shown: 0--10\% (top), 30--40\% (middle), 60--80\% (bottom).
The vertical bars on the data points indicate statistical uncertainties, while the shaded boxes indicate systematic uncertainties.
The widths of the boxes are not indicative of the bin size, and the points are shifted horizontally for better visibility.}
\label{fig:rdptr}
\end{figure}

In order to quantify the differences seen in Figure~\ref{fig:dptr}, ratios of the \Dptr\ distributions in \pbpb\ collisions to those measured in \pp\ collisions for $126 < \ptjet < 158$~\GeV\ and $200 < \ptjet < 251$~\GeV\ jets are presented in Figure~\ref{fig:rdptr}.
They are shown as a function of $r$ for different \pt\ and centrality selections.
In 0--10\% central collisions, \RDptr\ is greater than unity for $\rvar < 0.8$ for charged particles with \pT less than 4.0~\GeV\ for both \ptjet\ selections.
For these particles, the enhancement of yields in \pbpb\ collisions compared to those in \pp\ collisions grows with increasing \rvar\ up to approximately \mbox{$\rvar  = 0.3$}, with \RDptr\ reaching values up to two for 1.0~$< \pt <$~2.5~\GeV.
The value of \RDptr\ is approximately constant for \rvar\ in the interval \mbox{0.3--0.6} and decreases for \mbox{$\rvar > 0.6$}.
For charged particles with $\pt > 4.0$~\GeV, \RDptr\ shows a depletion outside the jet core for $r > 0.05$.
The magnitude of this depletion increases with increasing \rvar\ up to $r = 0.3$ and is approximately constant thereafter.
For 30--40\% mid-central collisions, the enhancement in the yield of particles with $\pt < 4.0$~\GeV\ has trends similar to those in the most central collisions, but the depletion of particles with $\pt > 4.0$~\GeV\ is not as strong.
For 60--80\% peripheral collisions, \RDptr\ has no significant \rvar\ dependence and the values of \RDptr\ are within approximately 50\% of unity.
 
The observed behavior inside the jet cone, $r < 0.4$, agrees with the measurement of the inclusive jet fragmentation functions~\cite{Aaboud:2017eww, Aaboud:2017bzv, Aaboud:2018hpb}, where yields of fragments with $\pt < 4$~\GeV\ are observed to be enhanced and yields of charged particles with intermediate \pT\ are suppressed in \PbPb\ collisions compared to those in \pp\ collisions.
The wake left in the medium from the passage of the jet compensates for some of the energy it loses \cite{Tachibana:2017syd, Casalderrey-Solana:2016jvj}.
The plateau and slight decrease seen in Figure~\ref{fig:rdptr} for the \RDptr\ distributions in central \pbpb\ collisions beyond $r = 0.6$ from the jet axis suggests that this response of the medium is smaller than that predicted in Ref.~\cite{Tachibana:2017syd}.

\begin{figure}
\includegraphics[width=1.\textwidth]{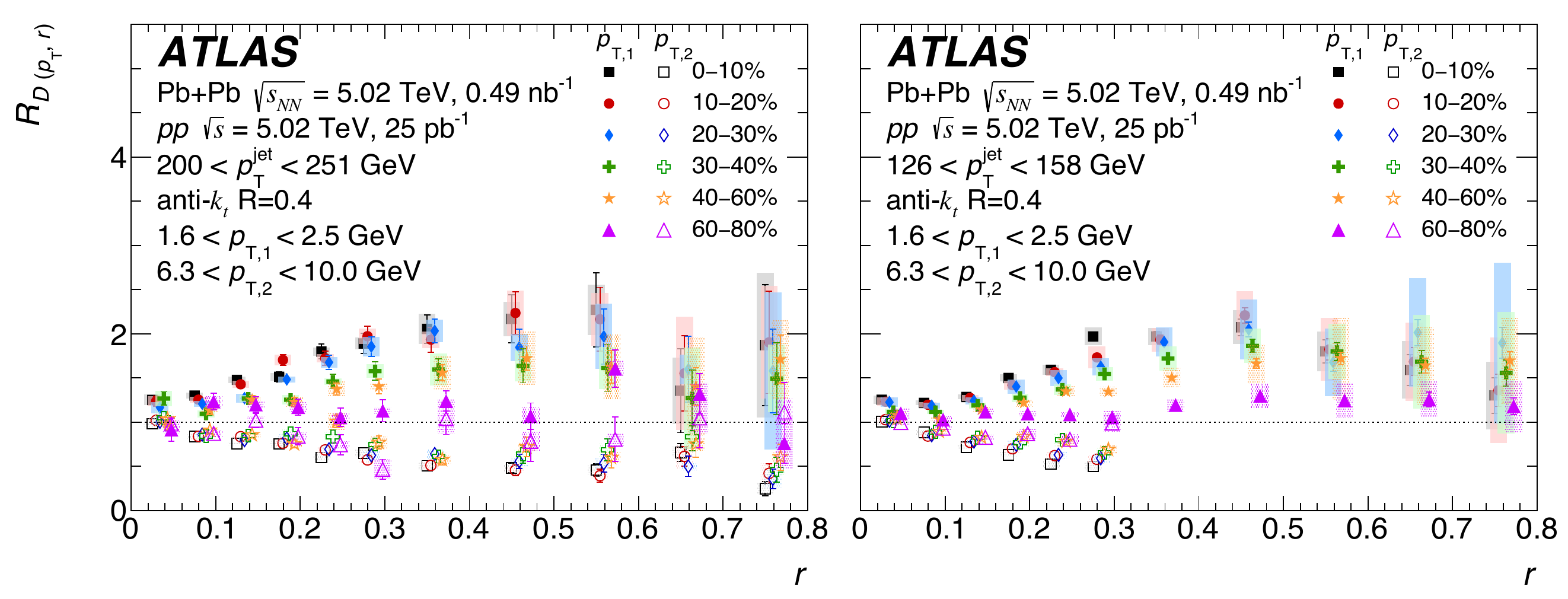}\hspace{0.0em}
\caption{The \RDptr\ distributions for \ptjet\ of 126--158~\GeV\ (left) and 200--251~\GeV\ (right) as a function of angular distance $r$ for two \pt\ selections, 1.6--2.5~\GeV\ (closed symbols) and 6.3--10.0~\GeV\ (open symbols), and six centrality intervals.
The vertical bars on the data points indicate statistical uncertainties, while the shaded boxes indicate systematic uncertainties.
The widths of the boxes are not indicative of the bin size, and the points are shifted horizontally for better visibility.}
\label{fig:centdep}
\end{figure}

The centrality dependence of \RDptr\ for two charged-particle \pt\ intervals (1.6--2.5~\GeV\ and \mbox{6.3--10.0~\GeV}) and two different \ptjet\ ranges (126--158~\GeV\ and 200--251~\GeV) is presented in Figure~\ref{fig:centdep}.
For both \ptjet\ selections, the magnitude of the excess for 1.6--2.5~\GeV\ charged particles increases for more-central events and for $\rvar < 0.3$.
The magnitude of the excess is approximately a factor of two in the most-central collisions for $\rvar >$~0.3.
A continuous centrality-dependent suppression of  yields of charged particles with $6.3 < \pt < 10.0$~\GeV\ is observed.
The magnitude of the modification decreases for more-peripheral collisions in both \pt\ intervals and \ptjet\ selections.
 
\begin{figure}
\includegraphics[width=1.\textwidth]{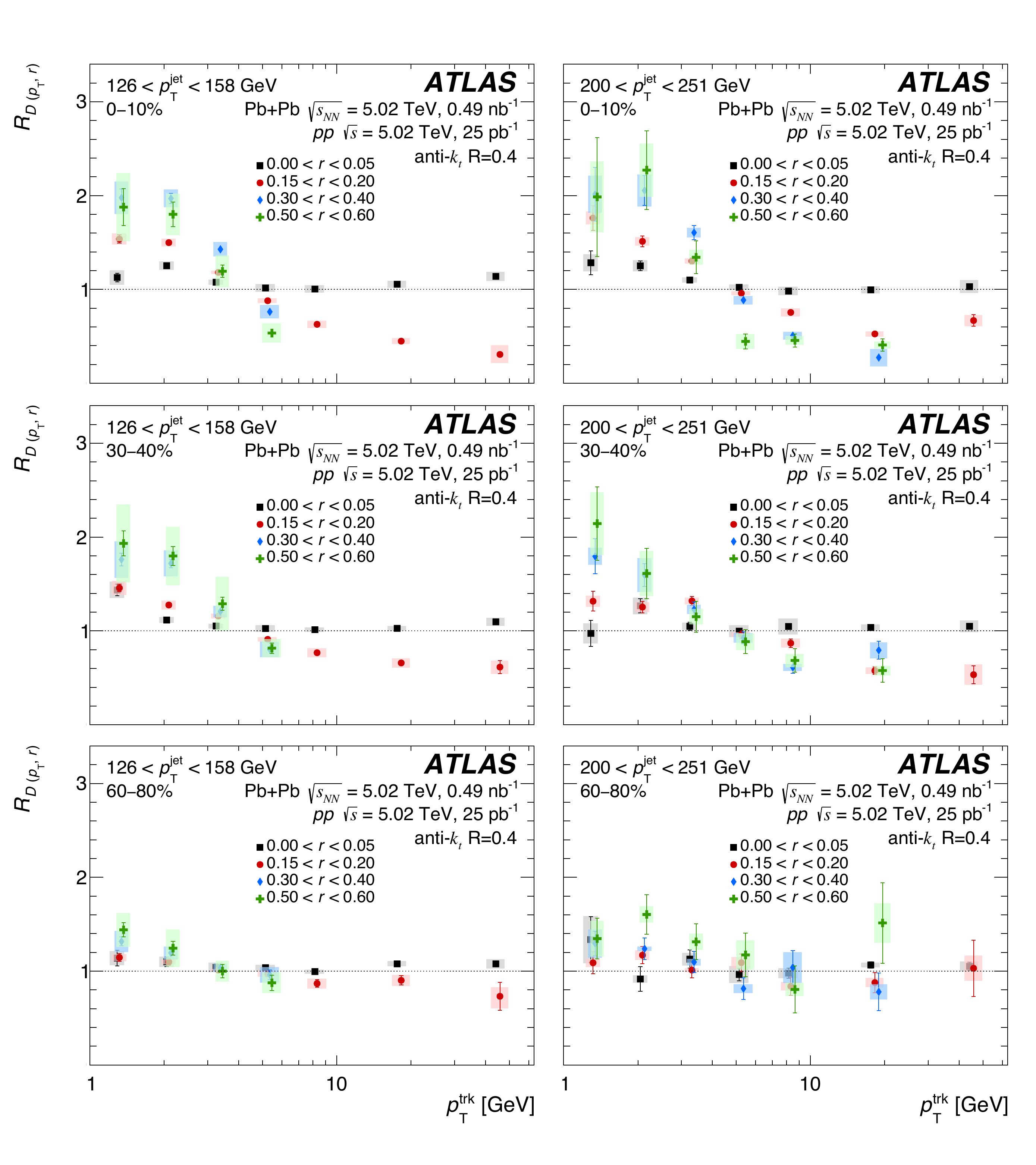}\hspace{0.0em}
\caption{\RDptr\ as a function of \pt\ for  0--10\% (top), 30--40\% (middle), and 60--80\% (bottom) \PbPb\ collisions in two different \ptjet\ selections: 126--158~\GeV\ (left) and 200--251~\GeV\ (right).
The different colors indicate different angular distances from the jet axis.
The vertical bars on the data points indicate statistical uncertainties, while the shaded boxes indicate systematic uncertainties.
The widths of the boxes are not indicative of the bin size, and the points are shifted horizontally for better visibility.}
\label{fig:pttrkdep}
\end{figure}

Figure~\ref{fig:pttrkdep} shows the charged-particle \pt\ dependence of \RDptr\ for selections in \rvar\  for 126--158~\GeV\ and 200--251~\GeV\ jets in the following centrality intervals: 0--10\%, 30--40\%, and 60--80\%.
Interestingly, there is no significant suppression of the yields in \pbpb\ collisions for $\rvar < 0.05$ at all measured \pt.
For larger \rvar\ values the yields are enhanced for charged particles with $\pt <$~4~\GeV\ and suppressed for higher-\pt\ charged particles in both the 0--10\% and 30--40\% centrality selections and both \ptjet\  ranges presented here.
The magnitude of the enhancement increases for decreasing \pt\ below 4~\GeV\ while the suppression is enhanced with increasing \pt\ for \mbox{4--10 \GeV}, after which it is approximately constant.
At fixed \pt\ the magnitude of the deviation from unity increases for larger distances from the jet axis.
In the 60--80\% peripheral collisions, the same trend is seen (but with smaller-magnitude modifications) for \mbox{$126 < \ptjet < 158$ \GeV}; for the higher-\ptjet\ selection the larger uncertainties do not allow a clear conclusion to be drawn for peripheral collisions.
 
The enhancement of the charged-particle yield in the kinematic region of \mbox{$\pt < 4$ \GeV} has two possible explanations.
First, gluon radiation from the hard-scattered parton as it propagates through the QGP would lead to extra soft particles \cite{Chien:2015vja, Kang:2017frl}.
Second, the interactions of a jet with the QGP and its hydrodynamic response could induce a wake that manifests itself as an enhancement in the number of low-\pt\ particles~\cite{Tachibana:2017syd}.
 
The observed modification at \mbox{$\pt > 4$ \GeV} can be explained on the basis of the larger expected energy loss of gluon-initiated jets, resulting in a relative enhancement in the number of quark jets in \pbpb\ collisions compared to \pp\ collisions at a given \ptjet\ value~\cite{Aaboud:2018hpb, Spousta:2015fca}.
Since gluon jets have a broader distribution of particle momentum transverse to the jet direction compared to quark-initiated jets \cite{OPAL:1995ab}, the enhanced quark-jet contribution could describe the narrowing of the particle distribution around the jet direction for particles with $\pt >$~4.0~\GeV\ observed here.

\begin{figure}
\makebox[\linewidth][c]{
\includegraphics[width=0.5\textwidth]{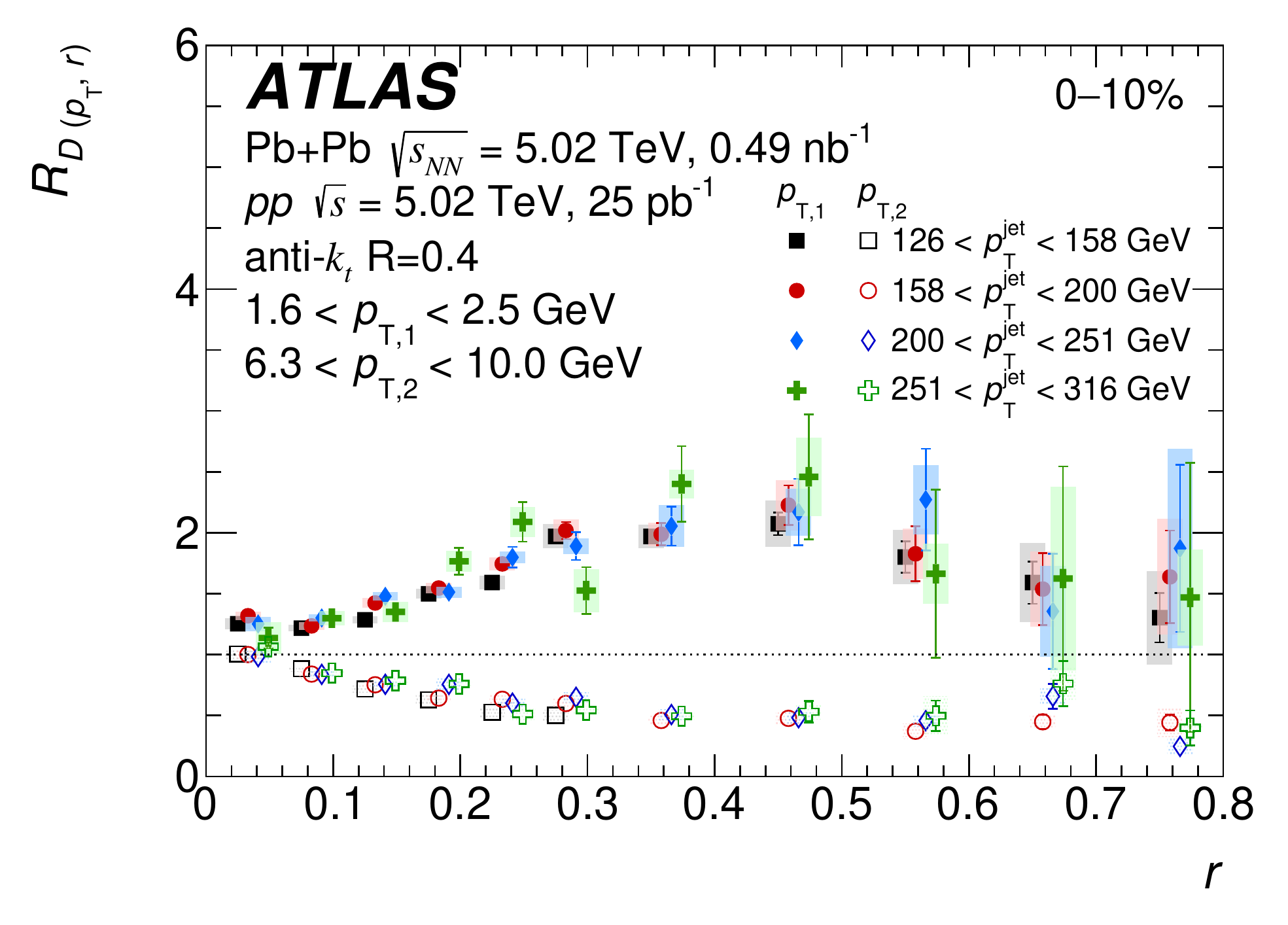}}
\caption{\RDptr\ as a function of \rvar\ for 0--10\% central collisions for charged particles with 1.6~$< \pt <$~2.5~\GeV\ (closed symbols) and 6.3~$< \pt <$10.0~\GeV\ (open symbols) for different \ptjet\ selections.
The vertical bars on the data points indicate statistical uncertainties, while the shaded boxes indicate systematic uncertainties.
The widths of the boxes are not indicative of the bin size, and the points are shifted horizontally for better visibility.}
\label{fig:ptjetdep}
\end{figure}

The \RDptr\ distributions for low- and high-\pt\ particles in the different \ptjet\ selections are directly overlaid in Figure~\ref{fig:ptjetdep}.
These distributions are for the 0--10\% most-central collisions, and show a hint of enhancement in \RDptr\ with increasing \ptjet\  for $r < 0.25$ for low-\pt\ charged particles.
No significant \ptjet\ dependence is seen at larger \rvar\ values, or for high-\pt\ charged particles at any \rvar.
This \ptjet\ dependence is further explored by defining an integral over the low-\pt\ excess and is discussed in Section~\ref{sec:discussion_int}.
 
\subsection{\DeltaDptr\ distributions}
\label{sec:delta_dptr}
In addition to the ratios of the \Dptr\ distributions, differences between the unfolded charged-particle yields are also evaluated as \DeltaDptr\ to quantify the modification in terms of the particle density.
 
\begin{figure}
\includegraphics[width=1.\textwidth]{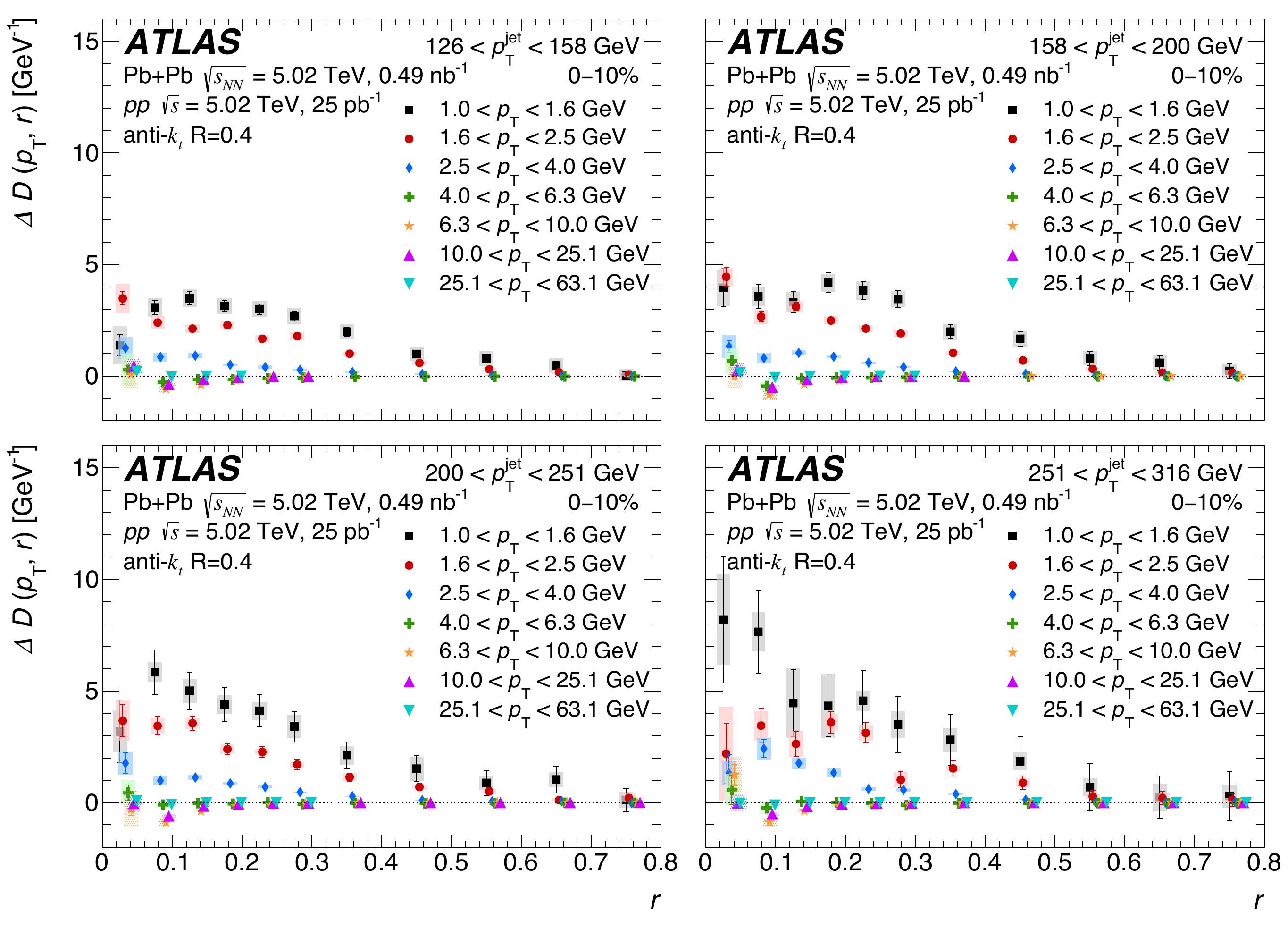}\hspace{0.0em}
\caption{\DeltaDptr\ as a function of \rvar\ in central collisions for all \pt\ ranges in four \ptjet\ selections: 126--158~\GeV, 158--200~\GeV, 200--251~\GeV, and 251--316~\GeV.
The vertical bars on the data points indicate statistical uncertainties, while the shaded boxes indicate systematic uncertainties.
The widths of the boxes are not indicative of the bin size, and the points are shifted horizontally for better visibility.}
\label{fig:deltadptr}
\end{figure}

These differences are presented as a function of $r$ for different \pt\ selections in 0--10\% central collisions in Figure~\ref{fig:deltadptr}.
These distributions show an excess in the charged-particle yield density for \pbpb\ collisions compared to \pp\ collisions for charged particles with $\pt <4.0$~\GeV.
This ranges from 0.5 to 4 particles per unit area per \GeV\ for charged particles with \mbox{$1 < \pt < 1.6$ GeV} in jets with \mbox{$126 < \ptjet < 158$ GeV} for 0--10\% central \pbpb\ collisions and increases with increasing \ptjet.
The largest excess for charged particles with $\pt <$~4.0~\GeV\ is within the jet cone.
For large \rvar\ values, the difference decreases but remains positive.
A depletion for higher-\pt\ particles of up to approximately 0.5 particles per unit area per \GeV\ is seen for 126--158~\GeV\ jets in 0--10\% central \pbpb\ collisions.
The magnitude of this depletion increases for higher \ptjet.
A minimum in the \DeltaDptr\ distributions for charged particles with \mbox{$ 4.0 < \pt <  25.1$}~\GeV\ at $0.05 < \rvar < 0.10$ is observed.
The magnitudes of the excesses and deficits discussed here depend on the selected charged-particle \pt.

\subsection{\pt\ integrated distributions}
\label{sec:discussion_int}
Motivated by similar studies of the enhancement of soft fragments in jet fragmentation functions in \pbpb\ compared to \pp\ collisions from Ref.~\cite{Aaboud:2018hpb}, the unfolded \Dptr\ distributions are integrated for charged particles with \pt\ < 4~\GeV\ to construct the quantities $\Theta(\rvar)$ and $P(\rvar)$ defined as:
 
\begin{align*}
\Theta(\rvar) &= \int_{1 \text{ GeV}}^{4 \text{ GeV}} \Dptr  d \pt \\
P(\rvar) &= \int_0^r \int_{1 \text{ GeV}}^{4 \text{ GeV}} D(\pt, r') d \pt d r' .
\end{align*}
The $\Theta(\rvar)$ values are integrated over the charged-particle \pt\ interval of 1--4~\GeV\ to provide a concise look at the \pt\ region of enhancement discussed above.
The $P(\rvar)$ values further add a running integral over \rvar\ and provide information about the jet shape.
Both of these quantities are compared between the \pp\ and \pbpb\ systems to give the following distributions:
 
\begin{align*}
\Delta_{\Theta(\rvar)} &= \Theta(\rvar)_{\mathrm{Pb+Pb}} - \Theta(\rvar)_{pp} \\
R_{\Theta(\rvar)} &= \frac{\Theta(\rvar)_{\mathrm{Pb+Pb}}}{\Theta(\rvar)_{pp}} \\
R_{P(\rvar)} &= \frac{P(\rvar)_{\mathrm{Pb+Pb}}}{P(\rvar)_{pp}}.
\end{align*}
These integrated quantities are intended to provide aggregate information about the variation with angular distance from the jet axis, magnitude, and \ptjet\ dependence of the low-\pt\ charged-particle excess discussed above.
The ratio quantities are useful for comparisons with other \pbpb\ measurements; $\Delta_{\Theta(\rvar)}$ is comparable to $\DeltaDptr$, but it is integrated over charged-particle \pt\ in the \mbox{1--4 \GeV} interval \cite{Aaboud:2018hpb}.
 
\begin{figure}
\makebox[\linewidth][c]{
\includegraphics[width=0.50\textwidth]{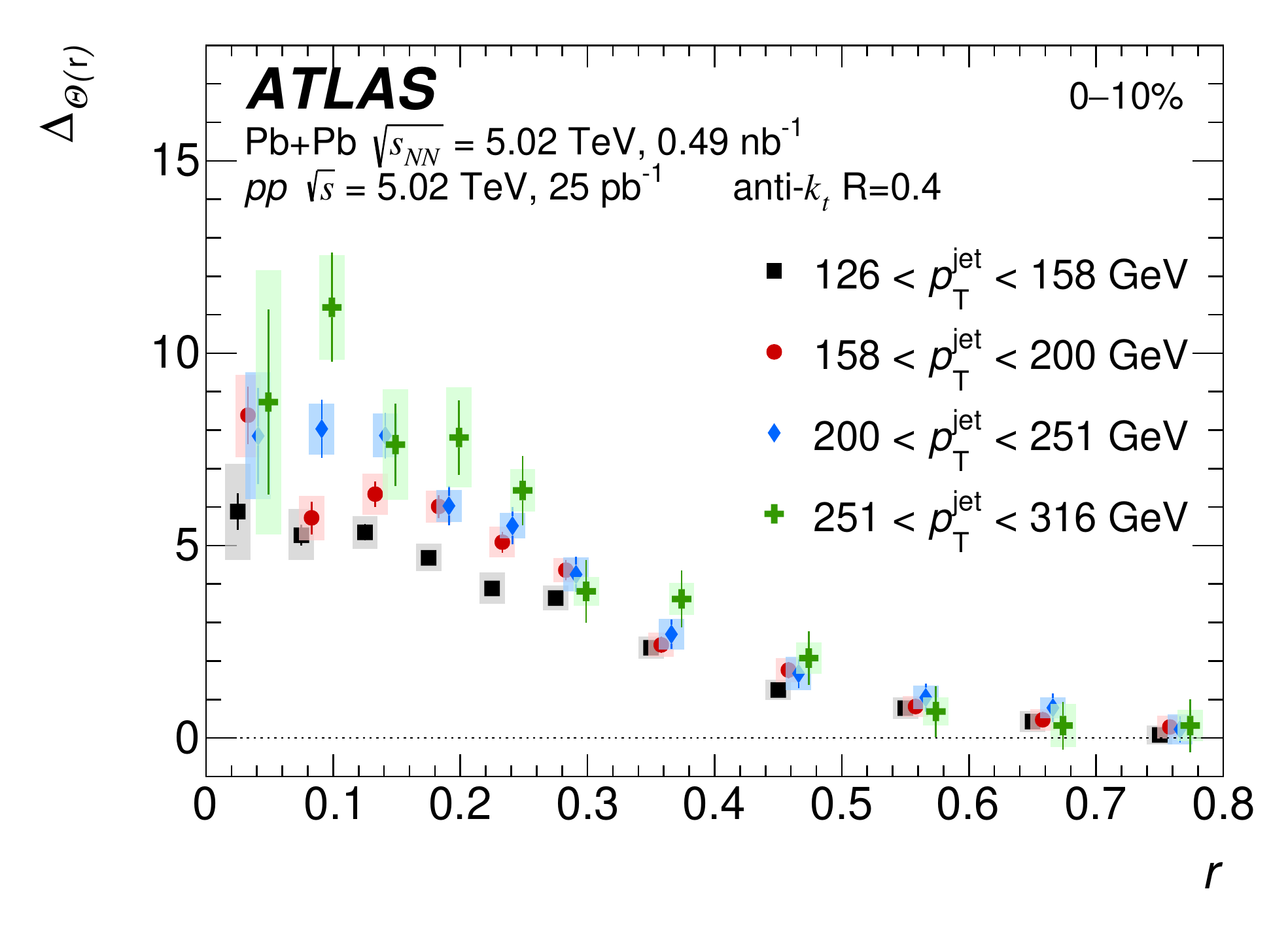}\hspace{0.0em}
\includegraphics[width=0.50\textwidth]{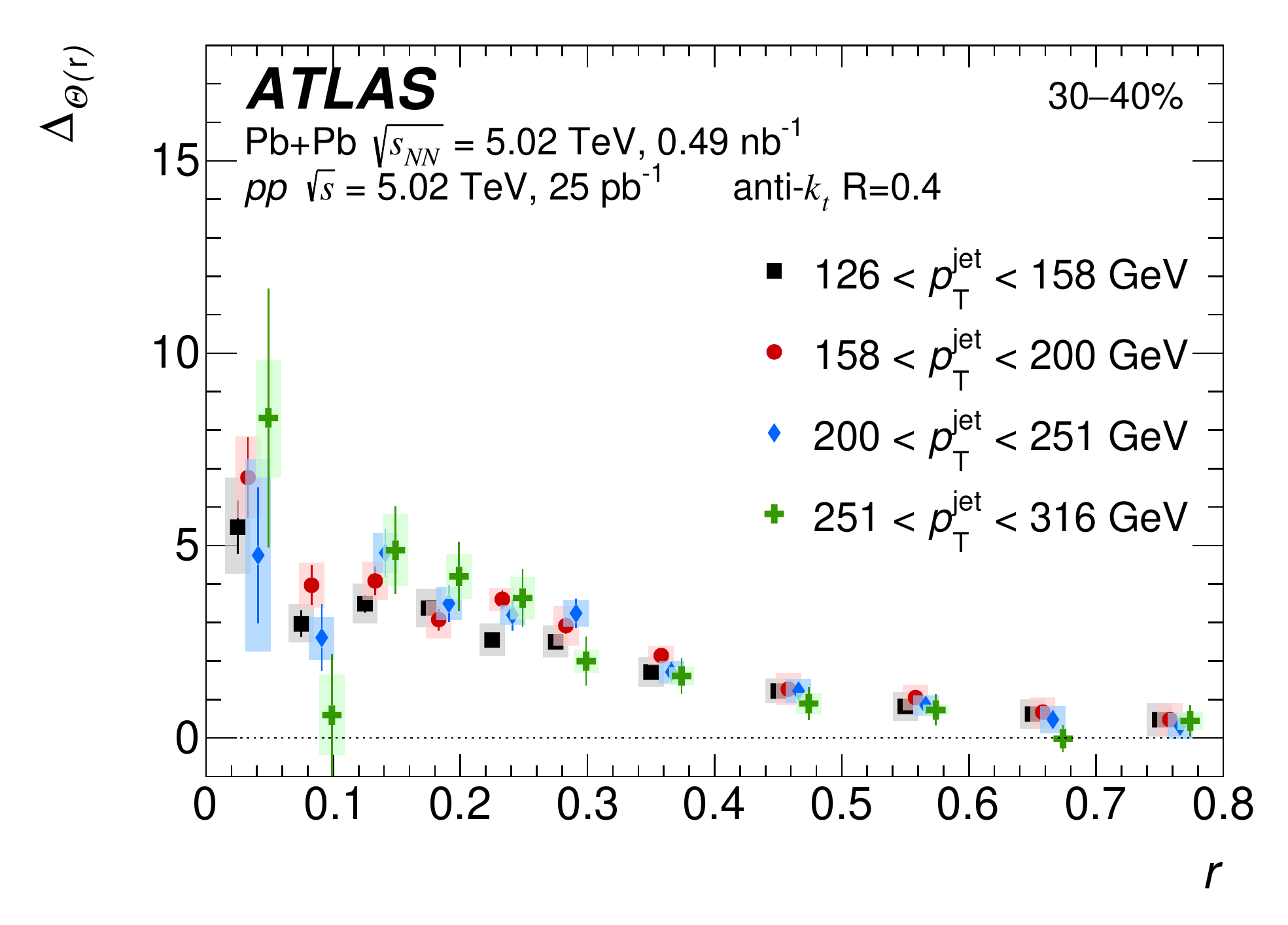}\hspace{0.0em}}
\centering
\includegraphics[width=0.50\textwidth]{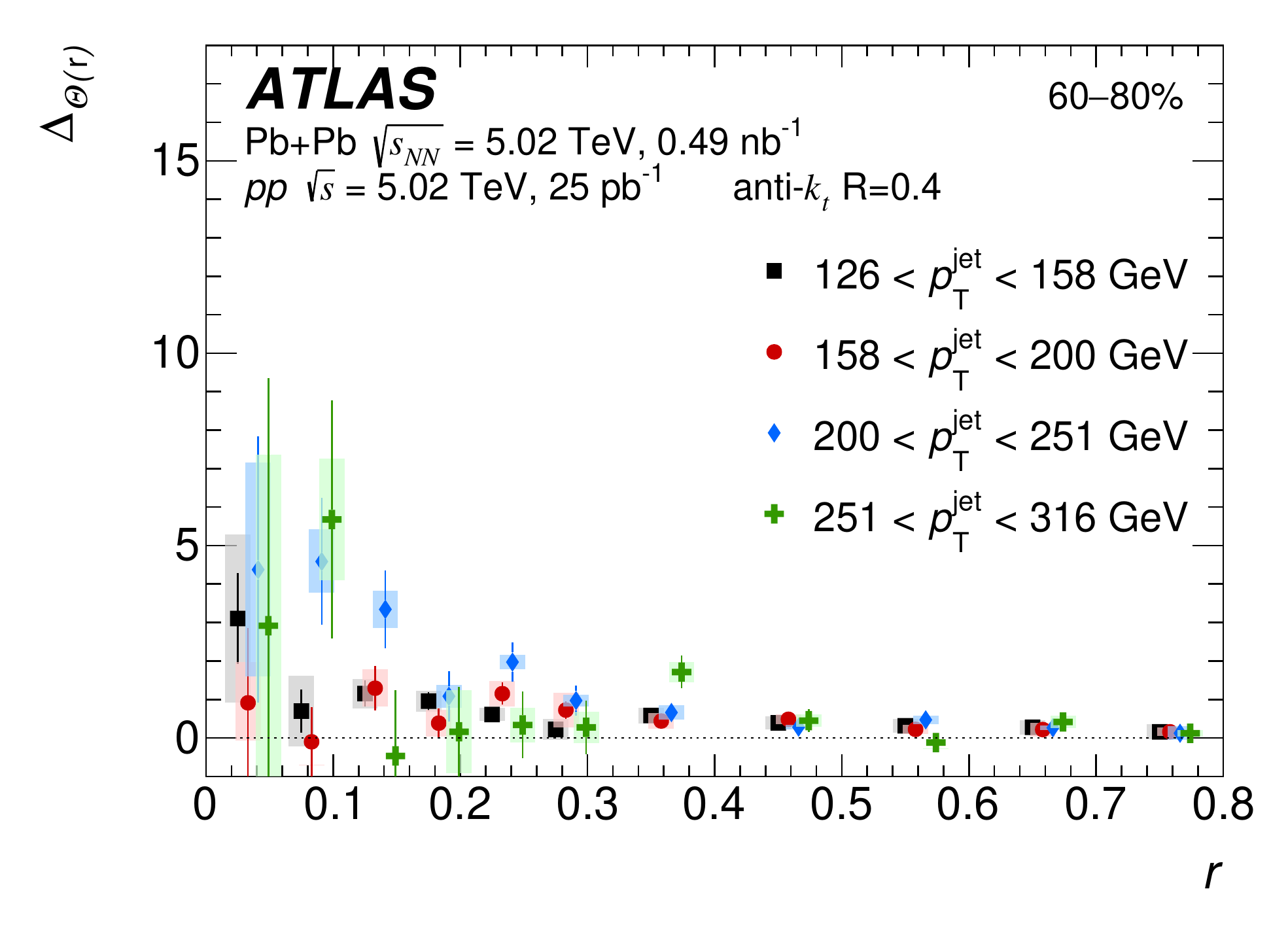}
\caption{\DeltaTheta\ as a function of \rvar\ for charged particles with \pt\ < 4~\GeV\  in four \ptjet\ selections: 126--158~\GeV, 158--200~\GeV, 200--251~\GeV, and 251--316~\GeV\ and three centrality selections: 0--10\% (top left), 30--40\% (top right) and 60--80\% (bottom).
The vertical bars on the data points indicate statistical uncertainties, while the shaded boxes indicate systematic uncertainties.
The widths of the boxes are not indicative of the bin size, and the points are shifted horizontally for better visibility.}
\label{fig:deltaPdeltaT}
\end{figure}

Figure~\ref{fig:deltaPdeltaT} shows the \DeltaTheta\ distributions as a function of \rvar\ for the following centrality intervals: 0--10\%, 30--40\%, and 60--80\%.
In the most-central collisions, a significant \ptjet\ dependence of \DeltaTheta\ is observed; for $\rvar <$~0.4 (particles within the jet cone) \DeltaTheta\ increases with increasing \ptjet.
The value of \DeltaTheta\ decreases in more-peripheral collisions, where its \ptjet\ dependence is also no longer significant.

\begin{figure}
\includegraphics[width=1.\textwidth]{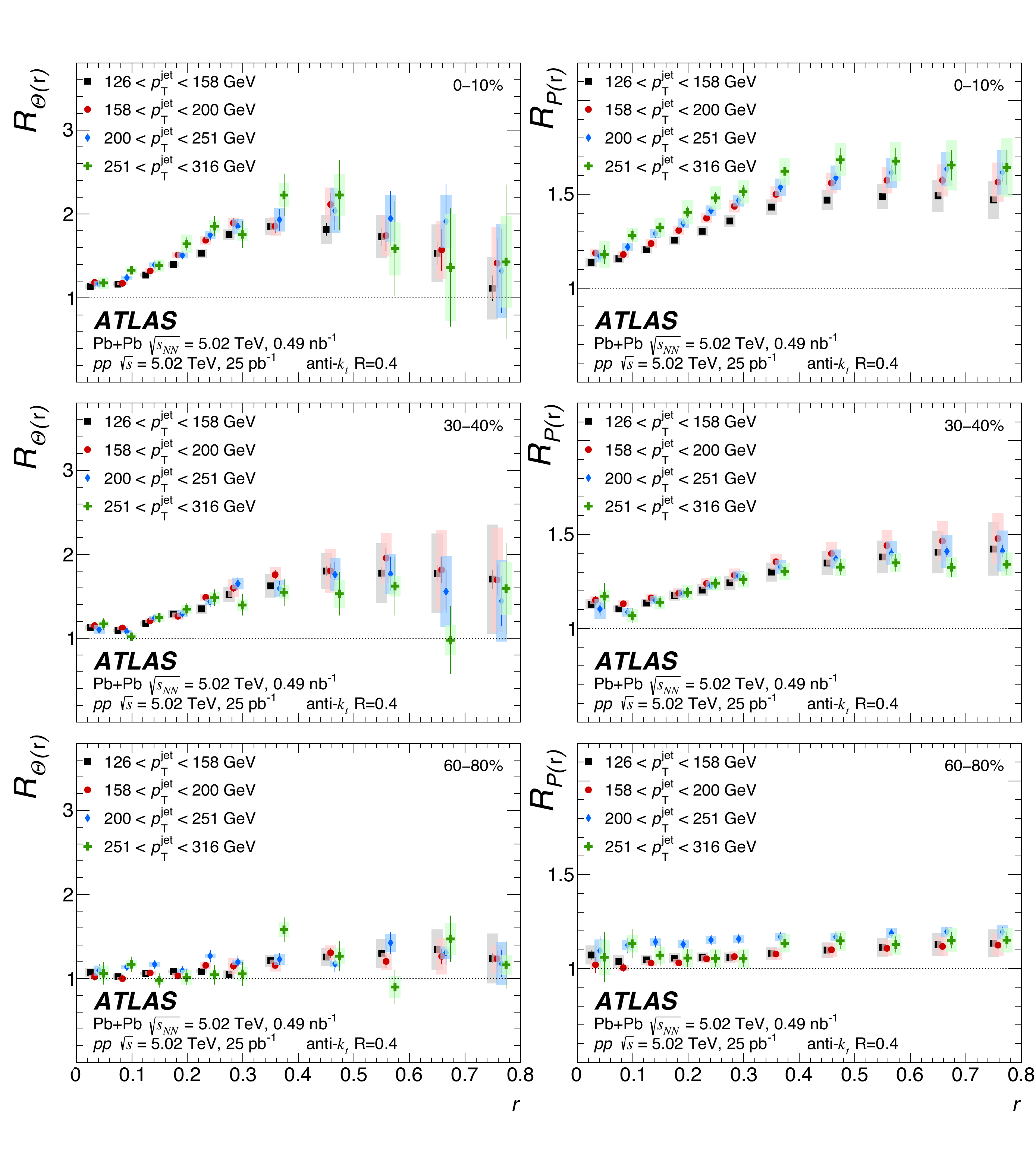}\hspace{0.0em}
\caption{\RTheta\ (left) and \RP\ (right) as a function of \rvar\ for charged particles with $\pt < 4$~\GeV\ ranges in four \ptjet\ selections: 126--158~\GeV, 158--200~\GeV, 200--251~\GeV, and 251--316~\GeV\ and three centrality selections: 0--10\% (top), 30--40\% (middle) and 60--80\% (bottom).
The vertical bars on the data points indicate statistical uncertainties, while the shaded boxes indicate systematic uncertainties.
The widths of the boxes are not indicative of the bin size, and the points are shifted horizontally for better visibility.}
\label{fig:RPRT}
\end{figure}

Figure~\ref{fig:RPRT} shows the \RTheta\ and \RP\ distributions as a function of \rvar\ for the 0--10\%, 30--40\%, and 60--80\% centrality intervals.
The \RTheta\ distributions in the most-central collisions show a maximum for $\rvar \sim 0.4$ and a flattening or a decrease for larger \rvar.
However, since \RTheta\ remains at or above unity for the full range of \rvar\ values presented, \RP\ shows no suppression with increasing \rvar\ over the entire measured range.
A significant \ptjet\ dependence is seen in the \RP\ distributions for the most-central \pbpb\ collisions.
A slow increase in \RP\ is clearly observed in 30--40\% central collisions.
In more-peripheral collisions the magnitude of the excess is reduced and the trends in \RTheta\ are less clear, although \RP\ is still seen to be above unity.
The flattening of the \RP\ distributions at large distances suggests that while wider jets have a softer fragmentation and contain more particles with less \pt\ in \pbpb\ collisions than in \pp\ \mbox{collisions \cite{Chesler2016, Hulcher:2017cpt}}, this effect plateaus for jets with radius larger than 0.6.

\FloatBarrier
\section{Summary}
\label{sec:summary}
 
This paper presents a measurement of the yields of charged particles, \Dptr, inside and around \RFour\ \antikt\ jets with $|\yjet| <$1.7 up to an angular distance of $r = 0.8$ from the jet axis, using the ATLAS detector at the LHC.
The yields are measured in intervals of \ptjet\ from 126 to 316~\GeV\ in \PbPb\ and \pp\ collisions at $\sqrtsnn=5.02$~\TeV\ as a function of charged-particle \pt\ and the angular distance \rvar\ between the jet axis and charged particle.
The integrated luminosities of the \PbPb\ and \pp\ data sets are 0.49~nb$^{-1}$ and 25~pb$^{-1}$, respectively.

The measurements show a broadening of the \Dptr\ distribution for low-\pt\ particles inside the jet in central \pbpb\ collisions compared to those in \pp\ collisions, while for higher-\pt\ particles the angular distributions are narrower in \pbpb\ collisions than in \pp\ collisions.
These modifications are centrality dependent and decrease for more-peripheral collisions.
The \PbPb-to-\pp\ ratio of the \Dptr\ distributions,~\RDptr,~is also examined.
The \RDptr\ distributions for charged particles with $\pt <$~4~\GeV\ are above unity and grow with increasing angular separation up to $r \sim0.3$, showing weak to no dependence on $r$ in the interval 0.3~$< \rvar <$~0.6 followed with a small decrease in the enhancement for 0.6~$< \rvar <$~0.8.
For charged particles with $\pt >$~4~\GeV, a suppression in \RDptr\ is observed and the size of the modification increases with increasing \rvar\ for 0.05 $ < \rvar < $~0.3 with no \rvar\ dependence for $r>0.3$.
For all charged-particle \pt\ values, the \RDptr\ values are greater than or equal to unity for $\rvar <$~0.05.
Between $0.1 < r < 0.25$, a statistically significant trend of increasing \RDptr\ with increasing \ptjet\ is observed for low-\pt\ particles.
No significant \ptjet\ dependence is seen for particles  with $\pt >$~4~\GeV.
 
This measurement provides information about the modification of the jet at large distances from the jet axis that can be used to constrain models of how the jet is modified by the presence of the quark--gluon plasma and how the quark--gluon plasma responds to the jet.
\section*{Acknowledgments}

We thank CERN for the very successful operation of the LHC, as well as the
support staff from our institutions without whom ATLAS could not be
operated efficiently.
 
We acknowledge the support of ANPCyT, Argentina; YerPhI, Armenia; ARC, Australia; BMWFW and FWF, Austria; ANAS, Azerbaijan; SSTC, Belarus; CNPq and FAPESP, Brazil; NSERC, NRC and CFI, Canada; CERN; CONICYT, Chile; CAS, MOST and NSFC, China; COLCIENCIAS, Colombia; MSMT CR, MPO CR and VSC CR, Czech Republic; DNRF and DNSRC, Denmark; IN2P3-CNRS and CEA-DRF/IRFU, France; SRNSFG, Georgia; BMBF, HGF and MPG, Germany; GSRT, Greece; RGC and Hong Kong SAR, China; ISF and Benoziyo Center, Israel; INFN, Italy; MEXT and JSPS, Japan; CNRST, Morocco; NWO, Netherlands; RCN, Norway; MNiSW and NCN, Poland; FCT, Portugal; MNE/IFA, Romania; MES of Russia and NRC KI, Russia Federation; JINR; MESTD, Serbia; MSSR, Slovakia; ARRS and MIZ\v{S}, Slovenia; DST/NRF, South Africa; MICINN, Spain; SRC and Wallenberg Foundation, Sweden; SERI, SNSF and Cantons of Bern and Geneva, Switzerland; MOST, Taiwan; TAEK, Turkey; STFC, United Kingdom; DOE and NSF, United States of America. In addition, individual groups and members have received support from BCKDF, CANARIE, Compute Canada and CRC, Canada; ERC, ERDF, Horizon 2020, Marie Sk{\l}odowska-Curie Actions and COST, European Union; Investissements d'Avenir Labex, Investissements d'Avenir Idex and ANR, France; DFG and AvH Foundation, Germany; Herakleitos, Thales and Aristeia programmes co-financed by EU-ESF and the Greek NSRF, Greece; BSF-NSF and GIF, Israel; La Caixa Banking Foundation, CERCA Programme Generalitat de Catalunya and PROMETEO and GenT Programmes Generalitat Valenciana, Spain; G\"{o}ran Gustafssons Stiftelse, Sweden; The Royal Society and Leverhulme Trust, United Kingdom.
 
The crucial computing support from all WLCG partners is acknowledged gratefully, in particular from CERN, the ATLAS Tier-1 facilities at TRIUMF (Canada), NDGF (Denmark, Norway, Sweden), CC-IN2P3 (France), KIT/GridKA (Germany), INFN-CNAF (Italy), NL-T1 (Netherlands), PIC (Spain), ASGC (Taiwan), RAL (UK) and BNL (USA), the Tier-2 facilities worldwide and large non-WLCG resource providers. Major contributors of computing resources are listed in Ref.~\cite{comp_Tier2s}.
 
\printbibliography

@article{Acharya:2019ssy,
      title          = "{Measurement of jet radial profiles in Pb$-$Pb collisions
                        at $\sqrtsnn = 2.76$ TeV}",
      author         = "{ALICE Collaboration}",
      journal        = "Phys. Lett.",
      volume         = "B796",
      year           = "2019",
      pages          = "204-219",
      doi            = "10.1016/j.physletb.2019.07.020",
      eprint         = "1904.13118",
      archivePrefix  = "arXiv",
      primaryClass   = "nucl-ex",
      reportNumber   = "CERN-EP-2019-079",
      SLACcitation   = "%%CITATION = ARXIV:1904.13118;%%"
}

@Article{Alioli:2010xa,
      author         = "Alioli, Simone and Hamilton, Keith and Nason, Paolo and
                        Oleari, Carlo and Re, Emanuele",
      title          = "{Jet pair production in POWHEG}",
      journal        = "JHEP",
      volume         = "04",
      year           = "2011",
      pages          = "081",
      doi            = "10.1007/JHEP04(2011)081",
      eprint         = "1012.3380",
      archivePrefix  = "arXiv",
      primaryClass   = "hep-ph",
      reportNumber   = "DESY-10-233, SFB-CPP-10-128, IPPP-10-100, DCPT-10-200,
                        MCNET-10-23",
      SLACcitation   = "%%CITATION = ARXIV:1012.3380;%%"
}

@Article{Alioli:2010xd,
      author         = "Alioli, Simone and Nason, Paolo and Oleari, Carlo and Re,
                        Emanuele",
      title          = "{A general framework for implementing NLO calculations in shower Monte Carlo programs: the POWHEG BOX}",
      journal        = "JHEP",
      volume         = "06",
      pages          = "043",
      doi            = "10.1007/JHEP06(2010)043",
      year           = "2010",
      eprint         = "1002.2581",
      archivePrefix  = "arXiv",
      primaryClass   = "hep-ph",
}

@Article{Frixione:2007vw, 
      author         = "Frixione, Stefano and Nason, Paolo and Oleari, Carlo",
      title          = "{Matching NLO QCD computations with Parton Shower
                        simulations: the POWHEG method}",
      journal        = "JHEP",
      volume         = "11",
      pages          = "070",
      doi            = "10.1088/1126-6708/2007/11/070",
      year           = "2007",
      eprint         = "0709.2092",
      archivePrefix  = "arXiv",
      primaryClass   = "hep-ph",
}

@article{Chesler2016,
      author         = "Chesler, Paul M. and Rajagopal, Krishna",
      title          = "{On the evolution of jet energy and opening angle in strongly coupled plasma}",
      journal        = "JHEP",
      volume         = "05",
      year           = "2016",
      pages          = "098",
      doi            = "10.1007/JHEP05(2016)098",
      eprint         = "1511.07567",
      archivePrefix  = "arXiv",
      primaryClass   = "hep-th",
      SLACcitation   = "%%CITATION = ARXIV:1511.07567;%%"
}

@article{Aad:2011sc,
      author         = "{ATLAS Collaboration}",
      title          = "{Measurement of the jet fragmentation function and
                        transverse profile in proton-proton collisions at a
                        center-of-mass energy of 7 TeV with the ATLAS detector}",
      collaboration  = "ATLAS",
      journal        = "Eur.~Phys.~J.~C",
      volume         = "71",
      year           = "2011",
      pages          = "1795",
      doi            = "10.1140/epjc/s10052-011-1795-y",
      eprint         = "1109.5816",
      archivePrefix  = "arXiv",
      primaryClass   = "hep-ex",
      reportNumber   = "CERN-PH-EP-2011-148",
      SLACcitation   = "%%CITATION = ARXIV:1109.5816;%%"
}

@article{Chatrchyan:2012mec,
      author         = "{CMS Collaboration}",
      title          = "{Shape, transverse size, and charged-hadron multiplicity of jets in pp collisions at $\sqrt{s}=7$ TeV}",
      collaboration  = "CMS",
      journal        = "JHEP",
      volume         = "06",
      year           = "2012",
      pages          = "160",
      doi            = "10.1007/JHEP06(2012)160",
      eprint         = "1204.3170",
      archivePrefix  = "arXiv",
      primaryClass   = "hep-ex",
      reportNumber   = "CMS-QCD-10-029, CERN-PH-EP-2012-079",
      SLACcitation   = "%%CITATION = ARXIV:1204.3170;%%"
}

@article{Aaboud:2016leb,
      author         = "{ATLAS Collaboration}",
      title          = "{Performance of the ATLAS trigger system in 2015}",
      collaboration  = "ATLAS",
      journal        = "Eur. Phys. J. C",
      volume         = "77",
      year           = "2017",
      number         = "5",
      pages          = "317",
      doi            = "10.1140/epjc/s10052-017-4852-3",
      eprint         = "1611.09661",
      archivePrefix  = "arXiv",
      primaryClass   = "hep-ex",
      reportNumber   = "CERN-EP-2016-241",
      SLACcitation   = "%%CITATION = ARXIV:1611.09661;%%"
}

@article{Aaboud:2017tke,
      author         = "{ATLAS Collaboration}",
      title          = "{Measurement of jet fragmentation in 5.02 TeV proton-lead
                        and proton-proton collisions with the ATLAS detector}",
      collaboration  = "ATLAS",
      journal        = "Nucl. Phys. A",
      volume         = "978",
      year           = "2018",
      pages          = "65",
      doi            = "10.1016/j.nuclphysa.2018.07.006",
      eprint         = "1706.02859",
      archivePrefix  = "arXiv",
      primaryClass   = "hep-ex",
      reportNumber   = "CERN-EP-2017-065",
      SLACcitation   = "%%CITATION = ARXIV:1706.02859;%%"
}

@article{Acharya:2018uvf,
      author         = "{ALICE Collaboration}",
      title          = "{Medium modification of the shape of small-radius jets in
                        central Pb-Pb collisions at $\sqrt{s_{\mathrm {NN}}} = 2.76 $ TeV}",
      collaboration  = "ALICE",
      journal        = "JHEP",
      volume         = "10",
      year           = "2018",
      pages          = "139",
      doi            = "10.1007/JHEP10(2018)139",
      eprint         = "1807.06854",
      archivePrefix  = "arXiv",
      primaryClass   = "nucl-ex",
      reportNumber   = "CERN-EP-2018-201",
      SLACcitation   = "%%CITATION = ARXIV:1807.06854;%%"
}

@article{2019167,
      author         = "{ATLAS Collaboration}",
      title          = "{Measurement of photon-jet transverse momentum
                        correlations in 5.02 TeV \pbpb\ and $pp$ collisions with
                        ATLAS}",
      collaboration  = "ATLAS",
      journal        = "Phys. Lett. B",
      volume         = "789",
      year           = "2019",
      pages          = "167-190",
      doi            = "10.1016/j.physletb.2018.12.023",
      eprint         = "1809.07280",
      archivePrefix  = "arXiv",
      primaryClass   = "nucl-ex",
      reportNumber   = "CERN-EP-2018-196",
      SLACcitation   = "%%CITATION = ARXIV:1809.07280;%%"
}

@article{Adam:2015ewa,
      author         = "{ALICE Collaboration}",
      title          = "{Measurement of jet suppression in central Pb-Pb collisions at $\sqrt{s_{\mathrm{NN}}} = 2.76$ TeV}",
      collaboration  = "ALICE",
      journal        = "Phys. Lett. B",
      volume         = "746",
      year           = "2015",
      pages          = "1-14",
      doi            = "10.1016/j.physletb.2015.04.039",
      eprint         = "1502.01689",
      archivePrefix  = "arXiv",
      primaryClass   = "nucl-ex",
      reportNumber   = "CERN-PH-EP-2015-023",
      SLACcitation   = "%%CITATION = ARXIV:1502.01689;%%"
}

@article{Ovanesyan:2011xy,
      author         = "Ovanesyan, Grigory and Vitev, Ivan",
      title          = "{An effective theory for jet propagation in dense QCD
                        matter: jet broadening and medium-induced bremsstrahlung}",
      journal        = "JHEP",
      volume         = "06",
      year           = "2011",
      pages          = "080",
      doi            = "10.1007/JHEP06(2011)080",
      eprint         = "1103.1074",
      archivePrefix  = "arXiv",
      primaryClass   = "hep-ph",
      SLACcitation   = "%%CITATION = ARXIV:1103.1074;%%"
}

@article{2019108,
      author         = "{ATLAS Collaboration}",
      title          = "{Measurement of the nuclear modification factor for
                        inclusive jets in \pbpb\ collisions at
                        $\sqrt{s_\mathrm{NN}}=5.02$ TeV with the ATLAS detector}",
      collaboration  = "ATLAS",
      journal        = "Phys. Lett. B",
      volume         = "790",
      year           = "2019",
      pages          = "108-128",
      doi            = "10.1016/j.physletb.2018.10.076",
      eprint         = "1805.05635",
      archivePrefix  = "arXiv",
      primaryClass   = "nucl-ex",
      reportNumber   = "CERN-EP-2018-105",
      SLACcitation   = "%%CITATION = ARXIV:1805.05635;%%"
}

@article{Aaboud:2018hpb,
      author         = "{ATLAS Collaboration}",
      title          = "{Measurement of jet fragmentation in \pbpb\ and $pp$
                        collisions at $\sqrt{s_{NN}} = 5.02$ TeV with the ATLAS
                        detector}",
      collaboration  = "ATLAS",
      journal        = "Phys. Rev. C",
      volume         = "98",
      year           = "2018",
      number         = "2",
      pages          = "024908",
      doi            = "10.1103/PhysRevC.98.024908",
      eprint         = "1805.05424",
      archivePrefix  = "arXiv",
      primaryClass   = "nucl-ex",
      reportNumber   = "CERN-EP-2018-096",
      SLACcitation   = "%%CITATION = ARXIV:1805.05424;%%"
}

@article{Sirunyan:2018jqr,
      author         = "{CMS Collaboration}",
      title          = "{Jet properties in PbPb and pp collisions at $\sqrt{{s}_{\mathrm{NN}}}=5.02$ TeV}",
      collaboration  = "CMS",
      journal        = "JHEP",
      volume         = "05",
      year           = "2018",
      pages          = "006",
      doi            = "10.1007/JHEP05(2018)006",
      eprint         = "1803.00042",
      archivePrefix  = "arXiv",
      primaryClass   = "nucl-ex",
      reportNumber   = "CMS-HIN-16-020, CERN-EP-2018-011",
      SLACcitation   = "%%CITATION = ARXIV:1803.00042;%%"
}

@article{Busza:2018rrf,
      author         = "Busza, Wit and Rajagopal, Krishna and van der Schee,
                        Wilke",
      title          = "{Heavy Ion Collisions: The Big Picture, and the Big
                        Questions}",
      journal        = "Ann. Rev. Nucl. Part. Sci.",
      volume         = "68",
      number         = "1",
      pages          = "339-376",
      doi            = "10.1146/annurev-nucl-101917-020852",
      year           = "2018",
      eprint         = "1802.04801",
      archivePrefix  = "arXiv",
      primaryClass   = "hep-ph",
      SLACcitation   = "%%CITATION = ARXIV:1802.04801;%%"
}

@article{Aaboud:2017bzv,
      author         = "{ATLAS Collaboration}",
      title          = "{Measurement of jet fragmentation in \pbpb\ and $pp$
                        collisions at $\sqrt{{s_\mathrm{NN}}} = 2.76$ TeV with the
                        ATLAS detector at the LHC}",
      collaboration  = "ATLAS",
      journal        = "Eur. Phys. J.~C",
      volume         = "77",
      year           = "2017",
      number         = "6",
      pages          = "379",
      doi            = "10.1140/epjc/s10052-017-4915-5",
      eprint         = "1702.00674",
      archivePrefix  = "arXiv",
      primaryClass   = "hep-ex",
      reportNumber   = "CERN-EP-2017-005",
      SLACcitation   = "%%CITATION = ARXIV:1702.00674;%%"
}

@article{OPAL:1995ab,
      author         = "{OPAL Collaboration}",
      title          = "{A model independent measurement of quark and gluon jet
                        properties and differences}",
      collaboration  = "OPAL",
      journal        = "Z. Phys. C",
      volume         = "68",
      year           = "1995",
      pages          = "179-202",
      reportNumber   = "CERN-PPE-95-075, CERN-PPE-95-75",
      doi            = "10.1007/BF01566667",
      SLACcitation   = "%%CITATION = ZEPYA,C68,179;%%"
}

@article{Aaboud:2017eww,
      author         = "{ATLAS Collaboration}",
      title          = "{Measurement of jet $p_{\mathrm{T}}$ correlations in
                        \pbpb\ and $pp$ collisions at $\sqrt{s_{\mathrm{NN}}}= 2.76$ TeV with the ATLAS detector}",
      collaboration  = "ATLAS",
      journal        = "Phys. Lett. B",
      volume         = "774",
      year           = "2017",
      pages          = "379-402",
      doi            = "10.1016/j.physletb.2017.09.078",
      eprint         = "1706.09363",
      archivePrefix  = "arXiv",
      primaryClass   = "hep-ex",
      reportNumber   = "CERN-EP-2017-054",
      SLACcitation   = "%%CITATION = ARXIV:1706.09363;%%"
}

@article{Khachatryan:2015lha,
      author         = "{CMS Collaboration}",
      title          = "{Measurement of transverse momentum relative to dijet
                        systems in PbPb and pp collisions at $
                        \sqrt{s_{\mathrm{NN}}}=2.76 $ TeV}",
      collaboration  = "CMS",
      journal        = "JHEP",
      volume         = "01",
      year           = "2016",
      pages          = "006",
      doi            = "10.1007/JHEP01(2016)006",
      eprint         = "1509.09029",
      archivePrefix  = "arXiv",
      primaryClass   = "nucl-ex",
      reportNumber   = "CMS-HIN-14-010, CERN-PH-EP-2015-260",
      SLACcitation   = "%%CITATION = ARXIV:1509.09029;%%"
}

@article{Aaboud:2018anc,
      author         = "{ATLAS Collaboration}",
      title          = "{Measurement of photon–jet transverse momentum
                        correlations in 5.02 TeV \pbpb\ and $pp$ collisions with
                        ATLAS}",
      collaboration  = "ATLAS",
      journal        = "Phys. Lett. B",
      volume         = "789",
      year           = "2019",
      pages          = "167-190",
      doi            = "10.1016/j.physletb.2018.12.023",
      eprint         = "1809.07280",
      archivePrefix  = "arXiv",
      primaryClass   = "nucl-ex",
      reportNumber   = "CERN-EP-2018-196",
      SLACcitation   = "%%CITATION = ARXIV:1809.07280;%%"
}

@article{Chatrchyan:2011sx,
      author         = "{CMS Collaboration}",
      title          = "{Observation and studies of jet quenching in PbPb
                        collisions at $\sqrt{s_{NN}}=2.76$ TeV}",
      collaboration  = "CMS",
      journal        = "Phys. Rev. C",
      volume         = "84",
      year           = "2011",
      pages          = "024906",
      doi            = "10.1103/PhysRevC.84.024906",
      eprint         = "1102.1957",
      archivePrefix  = "arXiv",
      primaryClass   = "nucl-ex",
      reportNumber   = "CERN-PH-EP-2011-001, CMS-HIN-10-004",
      SLACcitation   = "%%CITATION = ARXIV:1102.1957;%%"
}

@article{Aad:2010bu,
      author         = "{ATLAS Collaboration}",
      title          = "{Observation of a Centrality-Dependent Dijet Asymmetry in
                        Lead-Lead Collisions at $\sqrt{s_{NN}}=2.76$ TeV with the
                        ATLAS Detector at the LHC}",
      collaboration  = "ATLAS",
      journal        = "Phys. Rev. Lett.",
      volume         = "105",
      year           = "2010",
      pages          = "252303",
      doi            = "10.1103/PhysRevLett.105.252303",
      eprint         = "1011.6182",
      archivePrefix  = "arXiv",
      primaryClass   = "hep-ex",
      SLACcitation   = "%%CITATION = ARXIV:1011.6182;%%"
}

@article{Chatrchyan:2012gt,
      author         = "{CMS Collaboration}",
      title          = "{Studies of jet quenching using isolated-photon+jet
                        correlations in PbPb and $pp$ collisions at
                        $\sqrt{s_{NN}}=2.76$ TeV}",
      collaboration  = "CMS",
      journal        = "Phys. Lett. B",
      volume         = "718",
      year           = "2013",
      pages          = "773-794",
      doi            = "10.1016/j.physletb.2012.11.003",
      eprint         = "1205.0206",
      archivePrefix  = "arXiv",
      primaryClass   = "nucl-ex",
      reportNumber   = "CMS-HIN-11-010, CERN-PH-EP-2012-089",
      SLACcitation   = "%%CITATION = ARXIV:1205.0206;%%"
}

@article{Aaboud:2017jcu,
      author         = "{ATLAS Collaboration}",
      title          = "{Jet energy scale measurements and their systematic
                        uncertainties in proton-proton collisions at $\sqrt{s} =
                        13$ TeV with the ATLAS detector}",
      collaboration  = "ATLAS",
      journal        = "Phys. Rev. D",
      volume         = "96",
      year           = "2017",
      number         = "7",
      pages          = "072002",
      doi            = "10.1103/PhysRevD.96.072002",
      eprint         = "1703.09665",
      archivePrefix  = "arXiv",
      primaryClass   = "hep-ex",
      reportNumber   = "CERN-EP-2017-038",
      SLACcitation   = "%%CITATION = ARXIV:1703.09665;%%"
}

@article{Chien:2015vja,
      author         = "Chien, Yang-Ting and Emerman, Alexander and Kang,
                        Zhong-Bo and Ovanesyan, Grigory and Vitev, Ivan",
      title          = "{Jet quenching from QCD evolution}",
      journal        = "Phys. Rev. D",
      volume         = "93",
      year           = "2016",
      number         = "7",
      pages          = "074030",
      doi            = "10.1103/PhysRevD.93.074030",
      eprint         = "1509.02936",
      archivePrefix  = "arXiv",
      primaryClass   = "hep-ph",
      reportNumber   = "ACFI-T15-13",
      SLACcitation   = "%%CITATION = ARXIV:1509.02936;%%"
}

@article{Kang:2017frl,
      author         = "Kang, Zhong-Bo and Ringer, Felix and Vitev, Ivan",
      title          = "{Inclusive production of small radius jets in heavy-ion
                        collisions}",
      journal        = "Phys. Lett. B",
      volume         = "769",
      year           = "2017",
      pages          = "242-248",
      doi            = "10.1016/j.physletb.2017.03.067",
      eprint         = "1701.05839",
      archivePrefix  = "arXiv",
      primaryClass   = "hep-ph",
      SLACcitation   = "%%CITATION = ARXIV:1701.05839;%%"
}

@article{Wang:1991hta,
      title          = {\textsc{HIJING}: A Monte Carlo model for multiple jet production in $\mathrm{pp}$, $\mathrm{pA}$, and $\mathrm{AA}$ collisions},
      author         = {Wang, Xin-Nian and Gyulassy, Miklos},
      journal        = {Phys. Rev. D},
      volume         = {44},
      pages          = {3501--3516},
      numpages       = {0},
      year           = {1991},
      month          = {12},
      publisher      = {American Physical Society},
      doi            = {10.1103/PhysRevD.44.3501},
}

@article{Aaboud:2017all,
      author         = "{ATLAS Collaboration}",
      title          = "{Performance of the ATLAS track reconstruction algorithms
                        in dense environments in LHC Run 2}",
      collaboration  = "ATLAS",
      journal        = "Eur. Phys. J. C",
      volume         = "77",
      year           = "2017",
      number         = "10",
      pages          = "673",
      doi            = "10.1140/epjc/s10052-017-5225-7",
      eprint         = "1704.07983",
      archivePrefix  = "arXiv",
      primaryClass   = "hep-ex",
      reportNumber   = "CERN-EP-2017-045",
      SLACcitation   = "%%CITATION = ARXIV:1704.07983;%%"
}

@article{Chatrchyan:2013kwa,
      author         = "{CMS Collaboration}",
      title          = "{Modification of jet shapes in PbPb collisions at $\sqrt
                        {s_{\mathrm{NN}}} = 2.76$ TeV}",
      collaboration  = "CMS",
      journal        = "Phys.~Lett.~B",
      volume         = "730",
      year           = "2014",
      pages          = "243-263",
      doi            = "10.1016/j.physletb.2014.01.042",
      eprint         = "1310.0878",
      archivePrefix  = "arXiv",
      primaryClass   = "nucl-ex",
      reportNumber   = "CERN-PH-EP-2013-189, CMS-HIN-12-002",
      SLACcitation   = "%%CITATION = ARXIV:1310.0878;%%"
}

@article{Spousta:2015fca,
      author         = "Spousta, Martin and Cole, Brian",
      title          = "{Interpreting single jet measurements in \pbpb\
                        collisions at the LHC}",
      journal        = "Eur.~Phys.~J.~C ",
      volume         = "76",
      year           = "2016",
      number         = "2",
      pages          = "50",
      doi            = "10.1140/epjc/s10052-016-3896-0",
      eprint         = "1504.05169",
      archivePrefix  = "arXiv",
      primaryClass   = "hep-ph",
      SLACcitation   = "%%CITATION = ARXIV:1504.05169;%%"
}

@article{Abelev:2013kqa,
      author         = "{ALICE Collaboration}",
      title          = "{Measurement of charged jet suppression in Pb-Pb
                        collisions at $\sqrt{s_{NN}} = 2.76$ TeV}",
      collaboration  = "ALICE",
      journal        = "JHEP",
      volume         = "03",
      year           = "2014",
      pages          = "013",
      doi            = "10.1007/JHEP03(2014)013",
      eprint         = "1311.0633",
      archivePrefix  = "arXiv",
      primaryClass   = "nucl-ex",
      reportNumber   = "CERN-PH-EP-2013-205",
      SLACcitation   = "%%CITATION = ARXIV:1311.0633;%%"
}

@article{Nason:2004rx,
      author         = "Nason, Paolo",
      title          = "{A new method for combining NLO QCD with shower Monte
                        Carlo algorithms}",
      journal        = "JHEP",
      volume         = "11",
      year           = "2004",
      pages          = "040",
      doi            = "10.1088/1126-6708/2004/11/040",
      eprint         = "hep-ph/0409146",
      archivePrefix  = "arXiv",
      reportNumber   = "BICOCCA-FT-04-11",
      SLACcitation   = "%%CITATION = HEP-PH/0409146;%%"
}

@article{Sjostrand:2014zea,
      author         = "Sjöstrand, Torbjörn and Ask, Stefan and Christiansen,
                        Jesper R. and Corke, Richard and Desai, Nishita and Ilten,
                        Philip and Mrenna, Stephen and Prestel, Stefan and
                        Rasmussen, Christine O. and Skands, Peter Z.",
      title          = "{An introduction to PYTHIA 8.2}",
      journal        = "Comput. Phys. Commun.",
      volume         = "191",
      year           = "2015",
      pages          = "159-177",
      doi            = "10.1016/j.cpc.2015.01.024",
      eprint         = "1410.3012",
      archivePrefix  = "arXiv",
      primaryClass   = "hep-ph",
      reportNumber   = "LU-TP-14-36, MCNET-14-22, CERN-PH-TH-2014-190,
                        FERMILAB-PUB-14-316-CD, DESY-14-178, SLAC-PUB-16122",
      SLACcitation   = "%%CITATION = ARXIV:1410.3012;%%"
}

@article{Aaboud:2017tql,
      author         = "{ATLAS Collaboration}",
      title          = "{Measurement of longitudinal flow decorrelations in
                        \pbpb\ collisions at $\sqrt{s_\mathrm{NN}} = 2.76$ and 5.02
                        TeV with the ATLAS detector}",
      collaboration  = "ATLAS",
      journal        = "Eur. Phys. J.~C ",
      volume         = "78",
      year           = "2018",
      number         = "2",
      pages          = "142",
      doi            = "10.1140/epjc/s10052-018-5605-7",
      eprint         = "1709.02301",
      archivePrefix  = "arXiv",
      primaryClass   = "nucl-ex",
      reportNumber   = "CERN-EP-2017-191",
      SLACcitation   = "%%CITATION = ARXIV:1709.02301;%%"
}

@Booklet{ATLAS2014021,
      author         = "{ATLAS Collaboration}",
      title          = "{ATLAS Pythia 8 tunes to 7 TeV data}",
      howpublished        = "ATL-PHYS-PUB-2014-021",
      url            = "https://cds.cern.ch/record/1966419",
      year           = "2014",
}

@article{Aad:2008zzm,
      author         = "{ATLAS Collaboration}",
      title          = "{The ATLAS Experiment at the CERN Large Hadron Collider}",
      collaboration  = "ATLAS",
      journal        = "JINST",
      volume         = "3",
      pages          = "S08003",
      doi            = "10.1088/1748-0221/3/08/S08003",
      year           = "2008",
      SLACcitation   = "%%CITATION = JINST,3,S08003;%%",
}

@article{Roland:2014jsa,
      author         = "Roland, G. and \v{S}afa\v{r}\'{i}k, K. and Steinberg, P.",
      title          = "{Heavy-ion collisions at the LHC}",
      journal        = "Prog. Part. Nucl. Phys.",
      volume         = "77",
      year           = "2014",
      pages          = "70-127",
      doi            = "10.1016/j.ppnp.2014.05.001",
      SLACcitation   = "%%CITATION = PPNPD,77,70;%%"
}

@article{Hulcher:2017cpt,
      author         = "Hulcher, Zachary and Pablos, Daniel and Rajagopal,
                        Krishna",
      title          = "{Resolution effects in the hybrid strong/weak coupling model}",
      journal        = "JHEP",
      volume         = "03",
      year           = "2018",
      pages          = "010",
      doi            = "10.1007/JHEP03(2018)010",
      eprint         = "1707.05245",
      archivePrefix  = "arXiv",
      primaryClass   = "hep-ph",
      SLACcitation   = "%%CITATION = ARXIV:1707.05245;%%"
}

@article{Khachatryan:2016jfl,
      author         = "{CMS Collaboration}",
      title          = "{Measurement of inclusive jet cross sections in $pp$ and
                        PbPb collisions at $\sqrt{s_{NN}}= 2.76$ TeV}",
      collaboration  = "CMS",
      journal        = "Phys.~Rev.~C ",
      volume         = "96",
      year           = "2017",
      number         = "1",
      pages          = "015202",
      doi            = "10.1103/PhysRevC.96.015202",
      eprint         = "1609.05383",
      archivePrefix  = "arXiv",
      primaryClass   = "nucl-ex",
      reportNumber   = "CMS-HIN-13-005, CERN-EP-2016-217",
      SLACcitation   = "%%CITATION = ARXIV:1609.05383;%%"
}

@article{Ball:2012cx,
      author         = "Ball, Richard D. and others",
      title          = "{Parton distributions with LHC data}",
      journal        = "Nucl. Phys.~B ",
      volume         = "867",
      year           = "2013",
      pages          = "244-289",
      doi            = "10.1016/j.nuclphysb.2012.10.003",
      eprint         = "1207.1303",
      archivePrefix  = "arXiv",
      primaryClass   = "hep-ph",
      reportNumber   = "EDINBURGH-2012-08, IFUM-FT-997, FR-PHENO-2012-014,
                        RWTH-TTK-12-25, CERN-PH-TH-2012-037, SFB-CPP-12-47",
      SLACcitation   = "%%CITATION = ARXIV:1207.1303;%%"
}

@Article{Fastjet,
      author         = "Cacciari, Matteo and Salam, Gavin P. and Soyez, Gregory",
      title          = "{FastJet User Manual}",
      journal        = "Eur. Phys. J. C",
      volume         = "72",
      year           = "2012",
      pages          = "1896",
      doi            = "10.1140/epjc/s10052-012-1896-2",
      eprint         = "1111.6097",
      archivePrefix  = "arXiv",
      primaryClass   = "hep-ph",
      reportNumber   = "CERN-PH-TH-2011-297",
      SLACcitation   = "%%CITATION = ARXIV:1111.6097;%%"
}

@article{Cacciari:2008qp,
      author         = "Cacciari, Matteo and Salam, Gavin P. and Soyez, Gregory",
      title          = "{The \Antikt\ jet clustering algorithm}",
      journal        = "JHEP",
      volume         = "04",
      pages          = "063",
      doi            = "10.1088/1126-6708/2008/04/063",
      year           = "2008",
      eprint         = "0802.1189",
      archivePrefix  = "arXiv",
      primaryClass   = "hep-ph",
      reportNumber   = "LPTHE-07-03",
      SLACcitation   = "%%CITATION = ARXIV:0802.1189;%%",
}

@article{Chatrchyan:2014ava,
      author         = "{CMS Collaboration}",
      title          = "{Measurement of jet fragmentation in PbPb and $pp$
                        collisions at $\sqrt{s_{NN}}=2.76$ TeV}",
      collaboration  = "CMS",
      journal        = "Phys.~Rev.~C",
      volume         = "90",
      pages          = "024908",
      doi            = "10.1103/PhysRevC.90.024908",
      year           = "2014",
      eprint         = "1406.0932",
      archivePrefix  = "arXiv",
      primaryClass   = "nucl-ex",
      reportNumber   = "CMS-HIN-12-013, CERN-PH-EP-2014-100",
      SLACcitation   = "%%CITATION = ARXIV:1406.0932;%%",
}

@Article{Agostinelli:2002hh,
      author         = "Agostinelli, S. and others",
      title          = "{\textsc{Geant4}---a simulation toolkit}",
      collaboration  = "GEANT4",
      journal        = "Nucl. Instrum. Meth. A",
      volume         = "506",
      pages          = "250",
      doi            = "10.1016/S0168-9002(03)01368-8",
      year           = "2003",
      reportNumber   = "SLAC-PUB-9350, FERMILAB-PUB-03-339",
      SLACcitation   = "%%CITATION = NUIMA,A506,250;%%"
}

@Booklet{Adye:2011gm,
      author         = "Adye, Tim",
      title          = "{Unfolding algorithms and tests using RooUnfold}",
      year           = "2011",
      eprint         = "1105.1160",
      archivePrefix  = "arXiv",
      primaryClass   = "physics.data-an"
}

@article{Khachatryan:2016tfj,
      author         = "{CMS Collaboration}",
      title          = "{Decomposing transverse momentum balance contributions
                        for quenched jets in PbPb collisions at $
                        \sqrt{s_{\mathrm{NN}}}=2.76 $ TeV}",
      collaboration  = "CMS",
      journal        = "JHEP",
      volume         = "11",
      year           = "2016",
      pages          = "055",
      doi            = "10.1007/JHEP11(2016)055",
      eprint         = "1609.02466",
      archivePrefix  = "arXiv",
      primaryClass   = "nucl-ex",
      reportNumber   = "CMS-HIN-15-011, CERN-EP-2016-199",
      SLACcitation   = "%%CITATION = ARXIV:1609.02466;%%"
}

@article{Aad:2010ah,
      author         = "{ATLAS Collaboration}",
      title          = "{The ATLAS Simulation Infrastructure}",
      collaboration  = "ATLAS",
      journal        = "Eur.~Phys.~J.~C",
      volume         = "70",
      pages          = "823-874",
      doi            = "10.1140/epjc/s10052-010-1429-9",
      year           = "2010",
      eprint         = "1005.4568",
      archivePrefix  = "arXiv",
      primaryClass   = "physics.ins-det",
      SLACcitation   = "%%CITATION = ARXIV:1005.4568;%%",
}

@article{Aad:2011he,
      author         = "{ATLAS Collaboration}",
      collaboration  = "ATLAS",
      title          = "{Jet energy measurement with the ATLAS detector in
                        proton-proton collisions at $\sqrt{s}=7$ TeV}",
      journal        = "Eur. Phys. J.~C",
      volume         = "73",
      year           = "2013",
      pages          = "2304",
      doi            = "10.1140/epjc/s10052-013-2304-2",
      eprint         = "1112.6426",
      archivePrefix  = "arXiv",
      primaryClass   = "hep-ex",
      reportNumber   = "CERN-PH-EP-2011-191",
      SLACcitation   = "%%CITATION = ARXIV:1112.6426;%%"
}

@article{Aad:2012ag,
      author         = "{ATLAS Collaboration}",
      title          = "{Jet energy resolution in proton-proton collisions at
                        $\sqrt{s}=7$ TeV recorded in 2010 with the ATLAS
                        detector}",
      collaboration  = "ATLAS",
      journal        = "Eur.~Phys.~J.~C",
      volume         = "73",
      pages          = "2306",
      doi            = "10.1140/epjc/s10052-013-2306-0",
      year           = "2013",
      eprint         = "1210.6210",
      archivePrefix  = "arXiv",
      primaryClass   = "hep-ex",
      reportNumber   = "CERN-PH-EP-2012-191",
      SLACcitation   = "%%CITATION = ARXIV:1210.6210;%%",
}

@Booklet{ATLAS-TDR-2010-19,
    author         = "{ATLAS Collaboration}",
    title          = "{ATLAS Insertable B-Layer Technical Design Report}",
    howpublished   = "ATLAS-TDR-19",
    url            = "https://cds.cern.ch/record/1291633",
    year           = "2010",
    related        = "ATLAS-TDR-2010-19-add",
    relatedstring  = "Addendum:",
}

@article{Abbott:2018ikt,
      author         = "B. Abbott and others",
      title          = "{Production and integration of the ATLAS Insertable B-Layer}",
      journal        = "JINST",
      volume         = "13",
      year           = "2018",
      number         = "05",
      pages          = "T05008",
      doi            = "10.1088/1748-0221/13/05/t05008",
      eprint         = "1803.00844",
      archivePrefix  = "arXiv",
      primaryClass   = "physics.ins-det",
      SLACcitation   = "%%CITATION = ARXIV:1803.00844;%%"
}

@Booklet{HIjesnote,
      author         = "{ATLAS Collaboration}",
      title          = "{Jet energy scale and its uncertainty for jets
                       reconstructed using the ATLAS heavy ion jet algorithm}",
      howpublished   = "ATLAS-CONF-2015-016",
      year           = "2015",
      url            = "https://cds.cern.ch/record/2008677",
}

@Booklet{JERConfNote,
      author         = "{ATLAS Collaboration}",
      title          = "{Data-driven determination of the energy scale and resolution of jets reconstructed in the ATLAS calorimeters using dijet and multijet events at $\sqrt{s}=8$~TeV}",
      howpublished   = "{ATLAS-CONF-2015-017}",
      url            = "https://cds.cern.ch/record/2008678",
      year           = "2015",
}

@article{DAgostini:1994zf,
      author         = "D'Agostini, G.",
      title          = "{A multidimensional unfolding method based on Bayes' theorem}",
      journal        = "Nucl.~Instrum.~Meth.~A",
      volume         = "362",
      pages          = "487-498",
      doi            = "10.1016/0168-9002(95)00274-X",
      year           = "1995",
      reportNumber   = "DESY-94-099",
      SLACcitation   = "%%CITATION = NUIMA,A362,487;%%",
}

@article{Aad:2014wha,
      author         = "{ATLAS Collaboration}",
      title          = "{Measurement of inclusive jet charged-particle
                        fragmentation functions in \pbpb\ collisions at
                        $\sqrt{s_{NN}} = 2.76$ TeV with the ATLAS detector}",
      collaboration  = "ATLAS",
      journal        = "Phys.~Lett.~B",
      volume         = "739",
      pages          = "320-342",
      doi            = "10.1016/j.physletb.2014.10.065",
      year           = "2014",
      eprint         = "1406.2979",
      archivePrefix  = "arXiv",
      primaryClass   = "hep-ex",
      reportNumber   = "CERN-PH-EP-2014-101",
      SLACcitation   = "%%CITATION = ARXIV:1406.2979;%%",
}

@article{Aad:2014bxa,
      author         = "{ATLAS Collaboration}",
      title          = "{Measurements of the Nuclear Modification Factor for Jets
                        in \pbpb\ Collisions at $\sqrt{s_{\mathrm{NN}}}=2.76$ TeV
                        with the ATLAS Detector}",
      collaboration  = "ATLAS",
      journal        = "Phys.~Rev.~Lett.",
      volume         = "114",
      pages          = "072302",
      doi            = "10.1103/PhysRevLett.114.072302",
      year           = "2015",
      eprint         = "1411.2357",
      archivePrefix  = "arXiv",
      primaryClass   = "hep-ex",
      reportNumber   = "CERN-PH-EP-2014-172",
      SLACcitation   = "%%CITATION = ARXIV:1411.2357;%%",
}

@article{Blaizot:2014ula,
      author         = "Blaizot, Jean-Paul and Mehtar-Tani, Yacine and Torres,
                        Marcus A. C.",
      title          = "{Angular Structure of the In-Medium QCD Cascade}",
      journal        = "Phys. Rev. Lett.",
      volume         = "114",
      year           = "2015",
      number         = "22",
      pages          = "222002",
      doi            = "10.1103/PhysRevLett.114.222002",
      eprint         = "1407.0326",
      archivePrefix  = "arXiv",
      primaryClass   = "hep-ph",
      SLACcitation   = "%%CITATION = ARXIV:1407.0326;%%"
}

@article{Escobedo:2016jbm,
      author         = "Escobedo, Miguel A. and Iancu, Edmond",
      title          = "{Event-by-event fluctuations in the medium-induced jet
                        evolution}",
      journal        = "JHEP",
      volume         = "05",
      year           = "2016",
      pages          = "008",
      doi            = "10.1007/JHEP05(2016)008",
      eprint         = "1601.03629",
      archivePrefix  = "arXiv",
      primaryClass   = "hep-ph",
      SLACcitation   = "%%CITATION = ARXIV:1601.03629;%%"
}

@article{Casalderrey-Solana:2016jvj,
      author         = "Casalderrey-Solana, Jorge and Gulhan, Doga and Milhano,
                        Guilherme and Pablos, Daniel and Rajagopal, Krishna",
      title          = "{Angular structure of jet quenching within a hybrid
                        strong/weak coupling model}",
      journal        = "JHEP",
      volume         = "03",
      year           = "2017",
      pages          = "135",
      doi            = "10.1007/JHEP03(2017)135",
      eprint         = "1609.05842",
      archivePrefix  = "arXiv",
      primaryClass   = "hep-ph",
      reportNumber   = "ICCUB-16-033, CERN-TH-2016-201, MIT-CTP-4830",
      SLACcitation   = "%%CITATION = ARXIV:1609.05842;%%"
}

@article{Tachibana:2017syd,
      author         = "Tachibana, Yasuki and Chang, Ning-Bo and Qin, Guang-You",
      title          = "{Full jet in quark-gluon plasma with hydrodynamic medium
                        response}",
      journal        = "Phys. Rev. C",
      volume         = "95",
      year           = "2017",
      number         = "4",
      pages          = "044909",
      doi            = "10.1103/PhysRevC.95.044909",
      eprint         = "1701.07951",
      archivePrefix  = "arXiv",
      primaryClass   = "nucl-th",
      SLACcitation   = "%%CITATION = ARXIV:1701.07951;%%"
}

@article{Vitev:2008rz,
      author         = "Vitev, Ivan and Wicks, Simon and Zhang, Ben-Wei",
      title          = "{A theory of jet shapes and cross sections: from hadrons
                        to nuclei}",
      journal        = "JHEP",
      volume         = "11",
      year           = "2008",
      pages          = "093",
      doi            = "10.1088/1126-6708/2008/11/093",
      eprint         = "0810.2807",
      archivePrefix  = "arXiv",
      primaryClass   = "hep-ph",
      SLACcitation   = "%%CITATION = ARXIV:0810.2807;%%"
}

@article{Qin:2015srf,
      author         = "Qin, Guang-You and Wang, Xin-Nian",
      title          = "{Jet quenching in high-energy heavy-ion collisions}",
      journal        = "Int. J. Mod. Phys.~E",
      volume         = "24",
      year           = "2015",
      number         = "11",
      pages          = "1530014",
      doi            = "10.1142/S0218301315300143",
      eprint         = "1511.00790",
      archivePrefix  = "arXiv",
      primaryClass   = "hep-ph",
      SLACcitation   = "%%CITATION = ARXIV:1511.00790;%%"
}

@article{ATLAS:2012at,
      author         = "{ATLAS Collaboration}",
      title          = "{Measurement of the azimuthal anisotropy for charged
                        particle production in $\sqrt{s_{NN}}=2.76$ TeV lead-lead
                        collisions with the ATLAS detector}",
      collaboration  = "ATLAS",
      journal        = "Phys. Rev.~C",
      volume         = "86",
      year           = "2012",
      pages          = "014907",
      doi            = "10.1103/PhysRevC.86.014907",
      eprint         = "1203.3087",
      archivePrefix  = "arXiv",
      primaryClass   = "hep-ex",
      reportNumber   = "CERN-PH-EP-2012-035",
      SLACcitation   = "%%CITATION = ARXIV:1203.3087;%%"
}

\clearpage 
 
\begin{flushleft}
\hypersetup{urlcolor=black}
{\Large The ATLAS Collaboration}

\bigskip

\AtlasOrcid[0000-0002-6665-4934]{G.~Aad}$^\textrm{\scriptsize 102}$,    
\AtlasOrcid[0000-0002-5888-2734]{B.~Abbott}$^\textrm{\scriptsize 129}$,    
\AtlasOrcid{D.C.~Abbott}$^\textrm{\scriptsize 103}$,    
\AtlasOrcid[0000-0002-2788-3822]{A.~Abed~Abud}$^\textrm{\scriptsize 71a,71b}$,    
\AtlasOrcid[0000-0002-1002-1652]{K.~Abeling}$^\textrm{\scriptsize 53}$,    
\AtlasOrcid[0000-0002-2987-4006]{D.K.~Abhayasinghe}$^\textrm{\scriptsize 94}$,    
\AtlasOrcid[0000-0002-8496-9294]{S.H.~Abidi}$^\textrm{\scriptsize 167}$,    
\AtlasOrcid[0000-0002-8279-9324]{O.S.~AbouZeid}$^\textrm{\scriptsize 40}$,    
\AtlasOrcid{N.L.~Abraham}$^\textrm{\scriptsize 156}$,    
\AtlasOrcid[0000-0001-5329-6640]{H.~Abramowicz}$^\textrm{\scriptsize 161}$
% ...
\bigskip
\\

% ...
$^{40}$Niels Bohr Institute, University of Copenhagen, Copenhagen; Denmark.\\
% ...
$^{53}$II. Physikalisches Institut, Georg-August-Universit\"{a}t G\"ottingen, G\"ottingen; Germany.\\
% ...
$^{71}$$^{(a)}$INFN Sezione di Pavia;$^{(b)}$Dipartimento di Fisica, Universit\`a di Pavia, Pavia; Italy.\\
% ...
$^{94}$Department of Physics, Royal Holloway University of London, Egham; United Kingdom.\\
% ...
$^{102}$CPPM, Aix-Marseille Universit\'e, CNRS/IN2P3, Marseille; France.\\
$^{103}$Department of Physics, University of Massachusetts, Amherst MA; United States of America.\\
% ...
$^{129}$Homer L. Dodge Department of Physics and Astronomy, University of Oklahoma, Norman OK; United States of America.\\
% ...
$^{156}$Department of Physics and Astronomy, University of Sussex, Brighton; United Kingdom.\\
% ...
$^{161}$Raymond and Beverly Sackler School of Physics and Astronomy, Tel Aviv University, Tel Aviv; Israel.\\
% ...
$^{167}$Department of Physics, University of Toronto, Toronto ON; Canada.\\
% ...
$^{a}$ Also at Borough of Manhattan Community College, City University of New York, New York NY; United States of America.\\
$^{b}$ Also at CERN, Geneva; Switzerland.\\
% ...

\end{flushleft}


\end{document}